\documentclass[]{memtensor}

\usepackage[toc,page,header]{appendix}
\usepackage[utf8]{inputenc} 
\usepackage[T1]{fontenc}    
\usepackage{hyperref}       
\usepackage{url}            
\usepackage{booktabs}       
\usepackage{amsfonts}       
\usepackage{nicefrac}       
\usepackage{microtype}      
\usepackage{amsmath} 
\usepackage{etoolbox}
\usepackage{lipsum}  
\usepackage{minitoc}
\usepackage[table]{xcolor}  
\usepackage{tablefootnote}  
\usepackage{threeparttable}

\usepackage{listings}
\usepackage{titletoc}
\usepackage{marvosym}
\usepackage{tabularx}
\usepackage{longtable}
\usepackage{wrapfig}
\usepackage{multirow}
\usepackage{array}
\usepackage{arydshln}

\usepackage{algorithm}
\usepackage{algorithmic}
\usepackage{fontawesome5}

\title{%
  \titlefont%
  \fontsize{19pt}{20pt}\selectfont%
  \parbox{\textwidth}{%
    \centering
    \raisebox{-0.3em}{\includegraphics[height=1.3em]{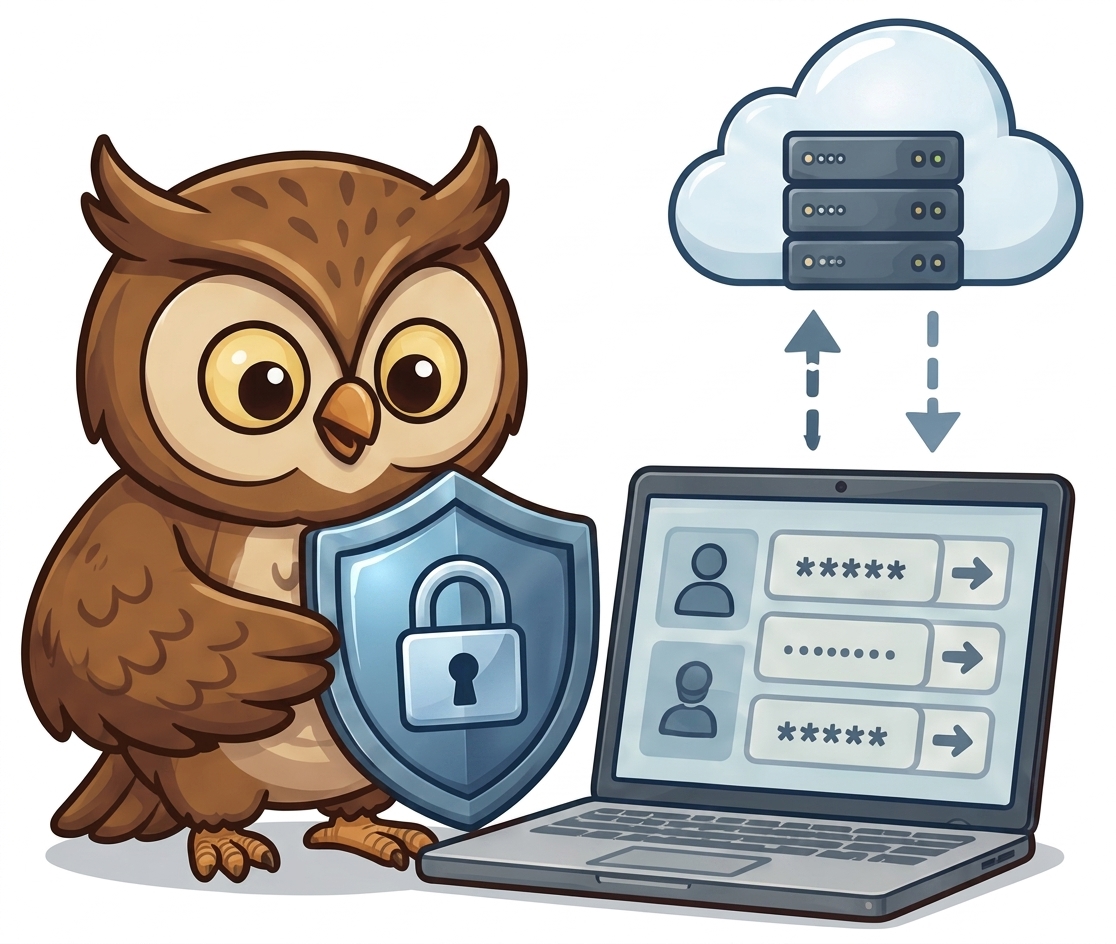}}\hspace{0.5em}%
    MemPrivacy: Privacy-Preserving Personalized Memory Management for Edge-Cloud Agents%
  }%
  \vspace{-1em}
}

\author[1]{Yining Chen$^{*,}$}
\author[1]{Jihao Zhao$^{*,}$}
\author[1]{Bo Tang}
\author[1,3]{Haofen Wang}
\author[2]{Yue Zhang}
\author[2]{Fei Huang}
\author[1]{Feiyu Xiong}
\author[1]{Zhiyu Li\textsuperscript{ \Letter{},}}

\affiliation[1]{MemTensor (Shanghai) Technology Co., Ltd. }
\affiliation[2]{HONOR Device Co., Ltd.}
\affiliation[3]{Tongji University}

\abstract{
As LLM-powered agents are increasingly deployed in edge-cloud environments, personalized memory has become a key enabler of long-term adaptation and user-centric interaction. However, cloud-assisted memory management exposes sensitive user information, while existing privacy protection methods typically rely on aggressive masking that removes task-relevant semantics and consequently degrades memory utility and personalization quality. To address this challenge, We propose MemPrivacy, which identifies privacy-sensitive spans on edge devices, replaces them with semantically structured type-aware placeholders for cloud-side memory processing, and restores the original values locally when needed. By decoupling privacy protection from semantic destruction, MemPrivacy minimizes sensitive data exposure while retaining the information required for effective memory formation and retrieval. We also construct MemPrivacy-Bench for systematic evaluation, a dataset covering 200 users and over 155k privacy instances, and introduce a four-level privacy taxonomy for configurable protection policies. Experiments show that MemPrivacy achieves strong performance in privacy information extraction, substantially surpassing strong general-purpose models such as GPT-5.2 and Gemini-3.1-Pro, while also reducing inference latency. Across multiple widely used memory systems, MemPrivacy limits utility loss to within 1.6\%, outperforming baseline masking strategies. Overall, MemPrivacy offers an effective balance between privacy protection and personalized memory utility for edge-cloud agents, enabling secure, practical, and user-transparent deployment.
}
\date{\today}
\correspondence{\email{lizy@memtensor.cn}}
\checkdata[Author Legend]{* Co-equal primary author, \Letter{} Corresponding authors}
\checkdata[Code]{\url{https://github.com/MemTensor/MemPrivacy}}
\checkdata[Model]{\url{https://huggingface.co/collections/IAAR-Shanghai/memprivacy}}

\begin{document}
\maketitle

\section{Introduction}

\begin{figure}[!htbp]
    \centering
    \includegraphics[width=\linewidth]{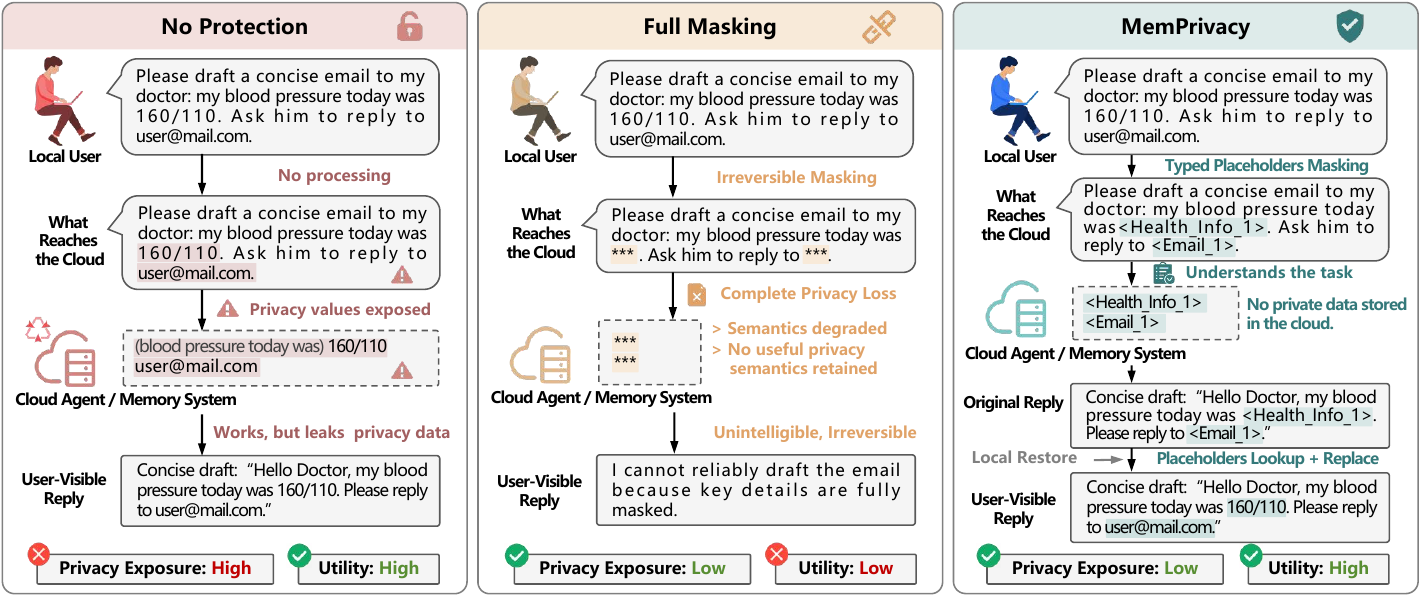}
    \caption{Comparison of privacy protection strategies for local-to-cloud agent interaction. Raw transmission exposes privacy data to the cloud, while full masking protects privacy at the cost of semantic utility. MemPrivacy instead uses semantic placeholders and local restoration to preserve both privacy and task effectiveness.}
    \label{fig:intro_comparison}
\end{figure}

With the rapid advancement of large language models (LLMs), intelligent agents are evolving from standalone text generators into memory-augmented systems capable of tool use, long-term adaptation, and personalized interaction~\cite{chen2025halumem,liu2026mememo}. In practical deployment, user interactions originate on edge devices, while computation-intensive reasoning and memory management are often offloaded to the cloud. This architecture makes personalized memory a key enabler of user-centric services, allowing agents to accumulate preferences, histories, and contextual knowledge over time. Recent memory systems, such as LongMem~\cite{wang2023augmenting} and Mem0~\cite{chhikara2025mem0}, as well as a growing number of conversational agents \cite{zhong2024memorybank,zhao2026inside}, have demonstrated the value of cloud-assisted memory for improving personalization quality and user experience. However, the more effectively agents leverage long-term memory for personalization, the more sensitive user information is exposed to cloud-side storage and processing.

This tension is particularly acute because memory management introduces a broader and more persistent privacy attack surface than one-shot cloud inference. User interactions naturally contain sensitive personally identifiable information (PII), including contact details, addresses, health conditions, financial information, and credentials. Once such content is transmitted in plaintext and incorporated into cloud logs, vector databases, or external memory stores, it may remain accessible throughout subsequent storage, retrieval, and reuse stages, creating opportunities for privacy leakage far beyond the original interaction. Prior studies have shown that multi-turn memory attacks can induce severe privacy violations with success rates up to 69\%~\cite{mireshghallah2025cimemories}, leakage attacks against memory systems can reach 75\% success~\cite{wang2025unveiling}, and indirect prompt injection can even manipulate agents into eliciting private information~\cite{cui2026vortexpia}. Beyond these technical threats, users often lack clear mental models of how cloud-based agents collect and reuse personal data, leading to anxiety, strategic self-censorship, and manual redaction behaviors that undermine utility~\cite{Shuning2025Perceptions}. Regulatory requirements such as the "right to be forgotten" further intensify the challenge, as deleting externally stored memories does not necessarily address information that has already been propagated through agent workflows or internalized by models~\cite{zhang2025right}. Privacy protection in cloud agent memory is thus not only necessary but urgent.

Existing countermeasures often face a fundamental trade-off, as illustrated in Figure~\ref{fig:intro_comparison}. Straightforward defenses such as full masking or redaction can prevent direct exposure of sensitive values, but they also remove critical semantic cues that support memory formation, retrieval, and downstream reasoning~\cite{mei2026accordingme,mukhopadhyay2025privacybench}. More principled techniques, including differential privacy, cryptographic protection, and anonymization, offer stronger privacy guarantees in specific settings, yet they are often difficult to integrate into interactive cloud inference and memory pipelines, or they incur substantial utility loss by obscuring task-relevant information. Moreover, users differ significantly in what they consider private and how strictly different categories of information should be protected, making one-size-fits-all protection strategies inadequate for personalized agents \cite{nissenbaum2004privacy}. These limitations expose a central challenge: how can an edge-cloud agent minimize sensitive data exposure during cloud-side memory processing while preserving the semantic structure necessary for accurate retrieval, long-term adaptation, and high-quality personalization?

To address this, we propose \textbf{MemPrivacy}, a privacy-preserving personalized memory management framework based on \textit{local reversible pseudonymization}. Instead of destroying sensitive content through coarse-grained masking, MemPrivacy performs privacy-sensitive span detection on edge devices and transforms raw private values into semantically structured, type-aware placeholders before cloud transmission. As shown in Figure~\ref{fig:intro_comparison}, a lightweight on-device MemPrivacy model identifies privacy spans, assigns each span a privacy type and protection level, and stores the original-to-placeholder mapping securely in a local database. The cloud-side memory system therefore receives semantically informative inputs that preserve the roles and relations needed for memory formation and retrieval, but it never directly observes the original sensitive values. When cloud processing is completed, MemPrivacy restores the protected values locally, enabling users to receive fluent and personalized responses without exposing raw private information to the cloud. To support configurable protection, MemPrivacy introduces a four-level privacy taxonomy that allows different protection policies to be applied according to user preferences and information sensitivity. We further construct \textbf{MemPrivacy-Bench}, a high-quality benchmark covering 200 users and more than 155K privacy instances, and use it to train lightweight MemPrivacy models ranging from 0.6B to 4B parameters for resource-constrained edge deployment. Extensive experiments show that MemPrivacy achieves stronger privacy-span extraction than powerful general-purpose LLMs, reduces inference latency, and preserves personalized memory utility substantially better than conventional masking strategies, thereby offering a practical and effective privacy-utility trade-off for edge-cloud agents.

Our main contributions are as follows:
\begin{itemize}
    \item We propose \textbf{MemPrivacy}, a privacy-preserving personalized memory management framework that reconciles privacy protection with cloud agent utility through typed placeholders and local restoration.
    \item We introduce a four-level privacy taxonomy that provides a standardized guideline for privacy identification and differential protection strategies.
    \item We construct \textbf{MemPrivacy-Bench}, a comprehensive dataset containing 200 users and 155k+ privacy instances, and release lightweight MemPrivacy models optimized for on-device deployment.
    \item Extensive evaluation across multiple models and memory systems confirms that MemPrivacy achieves state-of-the-art extraction performance and negligible utility loss compared to baseline methods.
\end{itemize}

\section{Related Work}
\subsection{Memory Operating Systems for LLM Agents}
To address the limited context windows of LLMs and enable continual adaptation across long-horizon interactions, recent work has increasingly treated memory as a core component of agent systems \cite{hu2025memory,zhao2026inside,kang2026memreader}. LongMem \cite{wang2023augmenting} augments frozen language models with an external memory bank to support long-range contextual modeling, while MemoryBank \cite{zhong2024memorybank} studies long-term conversational memory with mechanisms inspired by human forgetting. MemGPT \cite{packer2023memgpt} formulates memory management as OS-style virtual context handling, enabling agents to move information across memory tiers beyond the native context window. In practical agent frameworks, LangMem\footnote{\url{https://github.com/langchain-ai/langmem}} decouples memory primitives from background consolidation and persistence, and Mem0 \cite{chhikara2025mem0} proposes a scalable multi-level architecture for extracting and retrieving salient user information across sessions, and MemoBase\footnote{\url{https://github.com/memodb-io/memobase}} adopts a user-profile-centered design that combines structured profiles with time-aware event memories and buffered batch processing to support low-latency personalization. Moving beyond flat memory stores, A-Mem \cite{xu2025mem} organizes memory as an evolvable network with dynamic indexing and linking, whereas MemOS \cite{li2025memos_long} reconceptualizes memory as a first-class system resource and provides unified mechanisms for memory representation, organization, and lifecycle governance across heterogeneous memory forms. Overall, these studies indicate a shift from passive memory retrieval toward actively managed, structured memory systems for personalization, coherence, and long-term agent behavior.

\subsection{Privacy Protection for Long-Term Conversational Memory}
Existing research on LLMs privacy protection has spanned multiple stages, from training to deployment. However, its problem formulations remain noticeably misaligned with the protection needs of long-term memory dialogue systems. Some studies incorporate retrieval-augmented generation (RAG), cloud-edge collaborative inference, or prompt tuning into formal privacy frameworks based on differential privacy and cryptographic mechanisms, thereby reducing the risks of retrieval corpus leakage, query exposure, or fine-tuning data disclosure \cite{koga2024privacy,yao2025private,zhan2026prism,luo2025secp}. Yet such approaches either rely on noise injection and thus inevitably degrade the semantic fidelity of user facts, or primarily protect the retrieval or training process rather than the raw prompt content itself. As a result, they are unsuitable for scenarios that require the accurate preservation of genuine preferences, identity relations, and contextual constraints. Other studies focus on removing already absorbed sensitive knowledge from model parameters, including general unlearning frameworks for pretrained LLMs \cite{yao2024machine}, analyses of deletion targets under extraction attacks \cite{patil2023can}, and efficient unlearning methods based on LoRA and negative samples \cite{liu2025lune}. However, these works mainly address memorization during training rather than private content newly provided by users at inference time. Moreover, existing evidence suggests that ostensibly deleted knowledge may still be recoverable through intermediate-layer traces or paraphrasing attacks \cite{patil2023can}. In contrast, replacing private RAG contexts with fully synthetic data \cite{zeng2025mitigating}, or reducing sensitive corpus exposure through privacy-preserving vector databases and data processing frameworks \cite{huang2025dpf}, can indeed help mitigate the risk that models regurgitate pre-existing private data. Nevertheless, they still struggle to adequately cover a more realistic setting: users directly provide sensitive facts to a cloud-hosted model during inference and expect the system to retain and reuse them as long-term memory in subsequent interactions.

For this reason, recent studies have begun to reexamine privacy risks in long-term memory systems from multiple angles. MEXTRA directly reveals that the memory module itself has become an independent and high-risk surface for privacy exposure \cite{wang2025unveiling}. AirGapAgent advocates constraining the context accessible to agents under the principle of data minimization \cite{bagdasarian2024airgapagent}. Firewalls limits information flow and cross-module propagation in agentic networks through multilayer protective boundaries \cite{abdelnabi2025firewalls}. NeuroFilter enforces privacy guardrails using internal model activation signals \cite{das2026neurofilter}. Whistledown attempts to preserve conversational continuity through pseudonymization, local differential privacy, and caching \cite{mcmurray2025whistledown}. Meanwhile, PrivacyLens shows that a model’s awareness of privacy norms does not automatically translate into stable compliance during generation \cite{shao2024privacylens}. PrivacyBench further demonstrates that privacy risks in personalized dialogue can still be systematically evaluated and exposed \cite{mukhopadhyay2025privacybench}. User studies on RAG-based memory systems also indicate that users explicitly demand fine-grained control over memory, including inspectability, editability, deletability, and categorization \cite{zhang2025understanding}. Together, these findings suggest that relying solely on prompt engineering, post hoc filtering, or one-off refusals is insufficient to meet the privacy requirements of long-term memory systems. There is an urgent need to explore a new mechanism that proactively desensitizes inputs while preserving semantic utility, and that supports hierarchical policy configuration as well as consistent replacement across sessions.

\section{Problem Definition}
\label{subsec:formal_definition}

In interactions with edge-cloud agents, the central challenge of privacy protection is to design a mechanism that minimizes privacy leakage while preserving the capability of the agent and the user's personalized experience. We formulate this problem as a constrained optimization problem.

Let $X$ denote the user's raw input, and let $S = \{s_1, s_2, \dots, s_k\}$ be the set of privacy information contained in $X$. Let $\mathcal{C}$ denote the cloud-side agent, and let $\mathcal{M}$ denote its memory store or contextual state. In the ideal threat-free setting, the cloud agent directly receives the full plaintext input and produces the ideal response:
\begin{equation}
    Y_{\text{ideal}} = \mathcal{C}(X, \mathcal{M})
\end{equation}
This response represents the upper bound of utility, since the agent has access to all information.

To protect privacy, we introduce a local sanitization function $\mathcal{F}_{\text{san}}$ and a local restoration function $\mathcal{F}_{\text{res}}$. The raw input is first transformed into a safe sequence $X_{\text{safe}} = \mathcal{F}_{\text{san}}(X)$ before being sent to the cloud. Let $\mathcal{M}_{\text{safe}}$ denote the corresponding cloud-visible memory or context state under the same protection mechanism. The cloud model then performs inference on the sanitized input and produces an intermediate response $Y_{\text{safe}} = \mathcal{C}(X_{\text{safe}}, \mathcal{M}_{\text{safe}})$. Finally, the local device restores the response shown to the user:
\begin{equation}
    \hat{Y} = \mathcal{F}_{\text{res}}(Y_{\text{safe}})
\end{equation}

Based on this interaction process, we define two core metrics.

\textbf{1. Privacy Leakage Risk ($\mathcal{R}_{\text{priv}}$).}
This metric measures the probability that an attacker can recover any element of $S$ after observing $X_{\text{safe}}$, $Y_{\text{safe}}$, and $\mathcal{M}_{\text{safe}}$. Formally,
\begin{equation}
    \mathcal{R}_{\text{priv}}(\mathcal{F}_{\text{san}})
    =
    \Pr\!\left(
    \exists s \in S :
    s \in \mathcal{A}(X_{\text{safe}}, Y_{\text{safe}}, \mathcal{M}_{\text{safe}})
    \right)
\end{equation}
where $\mathcal{A}$ denotes an arbitrary privacy inference or memory extraction attack.

\textbf{2. Utility Loss ($\mathcal{L}_{\text{util}}$).}
This metric measures the gap between the final restored response $\hat{Y}$ and the ideal response $Y_{\text{ideal}}$, reflecting the degradation in both system utility and user experience. Let $\mathcal{U}$ denote an overall utility function. We define
\begin{equation}
    \mathcal{L}_{\text{util}}(\mathcal{F}_{\text{san}}, \mathcal{F}_{\text{res}})
    =
    \mathcal{U}(Y_{\text{ideal}}) - \mathcal{U}(\hat{Y})
\end{equation}

\textbf{Overall Objective.}
The goal of MemPrivacy is to find an optimal pair of local mapping functions $(\mathcal{F}_{\text{san}}^*, \mathcal{F}_{\text{res}}^*)$ that minimizes privacy leakage while keeping utility loss below a user-tolerable threshold:
\begin{equation}
\begin{gathered}
(\mathcal{F}_{\text{san}}^*, \mathcal{F}_{\text{res}}^*)
=
\arg\min_{\mathcal{F}_{\text{san}}, \mathcal{F}_{\text{res}}}
\mathcal{R}_{\text{priv}}(\mathcal{F}_{\text{san}}) \\
\text{s.t.}\ 
\mathcal{L}_{\text{util}}(\mathcal{F}_{\text{san}}, \mathcal{F}_{\text{res}}) \leq \epsilon
\end{gathered}
\end{equation}

This formulation clarifies the goal of this work: to minimize privacy leakage while limiting utility loss, thereby providing effective privacy protection without perceptibly degrading the user experience.

\section{The MemPrivacy Framework}

\subsection{Overview Architecture}

\label{subsec:workflow}

\begin{figure}[!htbp]
    \centering
    \includegraphics[width=\linewidth]{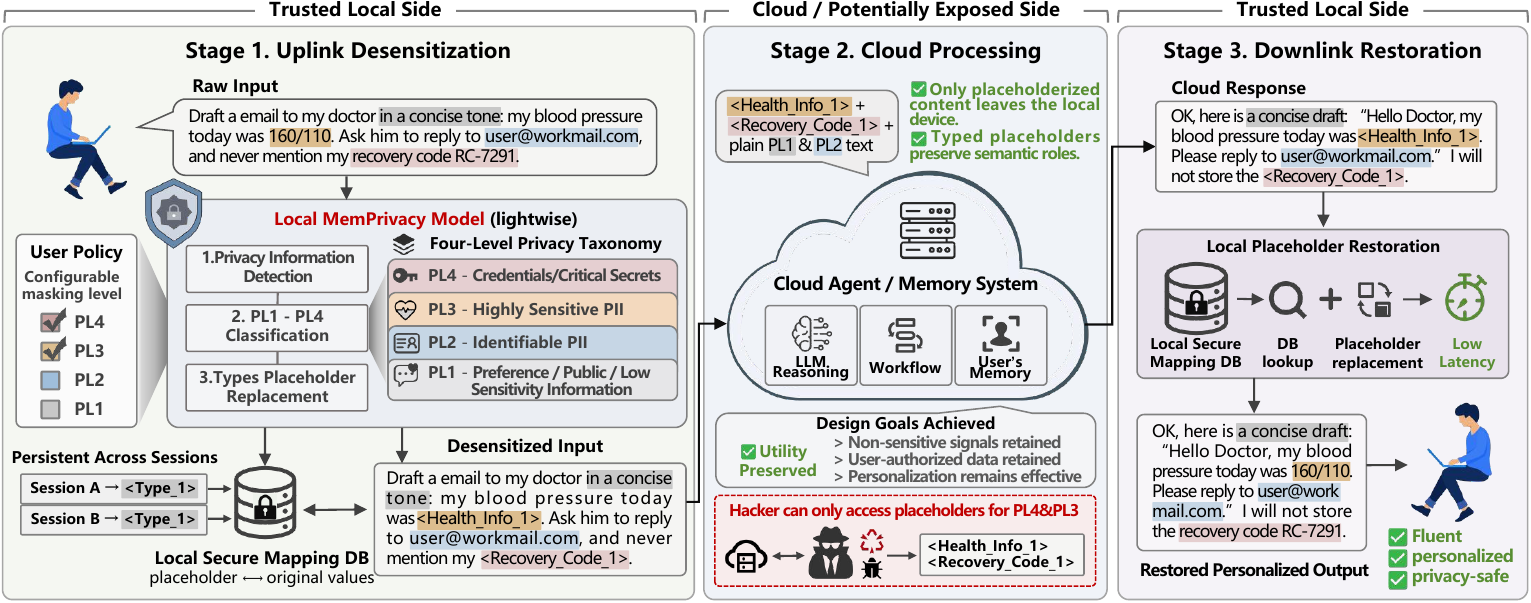}
    \caption{Overview of MemPrivacy, a closed-loop framework for privacy-preserving cloud agent memory. User inputs are desensitized on device according to privacy levels and types, processed in the cloud with typed placeholders, and finally restored locally for transparent user experience.}
    \label{fig:memprivacy_framework}
\end{figure}

As shown in Figure~\ref{fig:memprivacy_framework}, MemPrivacy follows a three-stage lifecycle, forming a fully closed-loop and user-transparent privacy protection framework for edge-cloud agents.

\textbf{Stage 1: Uplink Desensitization.} When a user issues a request on the local device, the input is processed before leaving the device. A lightweight on-device \textbf{MemPrivacy Model} first identifies privacy spans in the input and produces a structured output for each span, consisting of the original span text, its privacy level, and its privacy type under the PL2--PL4 taxonomy. Based on the detected privacy type, the system replaces protected spans with semantic typed placeholders, e.g., \texttt{<EMAIL\_1>}, while spans of the same type are distinguished by incremental indices. To support long-term memory, the mapping between original values and placeholders is securely stored in a local database, enabling consistent restoration across sessions. Users can also configure the masking threshold, e.g., masking only PL3 and PL4, to achieve fine-grained control over the privacy--utility trade-off.

\textbf{Stage 2: Cloud Processing.} The desensitized input is then sent to the cloud for task execution and memory operations. Because typed placeholders preserve semantic type information, the cloud model can still perform accurate language understanding and reasoning. At the same time, the high-precision privacy recognition of MemPrivacy model ensures that non-sensitive personalized signals, as well as user-authorized privacy levels, remain available to the cloud, avoiding the semantic damage caused by over-masking. As a result, the cloud system can maintain personalization and utility, while any leaked cloud-side content reveals only semantically typed placeholders rather than usable private values, achieving architecture-level privacy isolation.

\textbf{Stage 3: Downlink Restoration.} To preserve user experience, MemPrivacy provides a low-latency local restoration mechanism. After the cloud returns a response that may contain placeholders, the local system queries the local database and replaces each placeholder with its original value. Since this process only involves lightweight database lookup and string substitution, its overhead is negligible. The user therefore sees a fluent and fully personalized response, while privacy protection remains entirely transparent during interaction.

Based on the above architectural design, the end-to-end execution flow of our framework can be formalized as Algorithm~\ref{alg:memprivacy_framework}. It presents how the proposed modules are coordinated in practice, from local privacy identification to cloud inference and final local recovery.

\subsection{Four-Level Privacy Taxonomy (PL1--PL4)}
\label{Four-Level Privacy Taxonomy}
We introduce a \textbf{four-level privacy taxonomy} for privacy identification and differential protection. It organizes privacy-relevant content by identifiability, expected harm, and operational exploitability. Concrete examples and the corresponding prompts are provided in the appendix.

At the lowest end, PL1 functions primarily as an exclusion class. It covers generic preferences, habits, stylistic choices, and non-diagnostic self-descriptions that do not, by themselves, identify a specific natural person and do not ordinarily create substantial downstream harm. This boundary is consistent with contextual theories of privacy, which emphasize that privacy risk depends not merely on whether information is personal in an everyday sense, but on whether its collection or dissemination becomes identifying, inappropriate, or harmful in context \cite{solove2023data}. It is also consistent with empirical work showing that perceived sensitivity varies with public availability, context of use, and identifiability, rather than being uniform across all self-related information~\cite{nissenbaum2004privacy}. 

PL2 captures information that can identify, locate, or stably trace a natural person, either directly or when linked with reasonably available auxiliary information. This notion closely follows mainstream legal definitions of personal data, which center on whether a person is identified or identifiable, including through indirect reference to identifiers such as names, identification numbers, location data, and online identifiers \cite{gellert2021personal,erdos2022identification}. The same logic is reinforced by the re-identification literature, as both k-anonymity and subsequent critiques of anonymization show that data fields that appear innocuous in isolation can become identifying once linked across datasets \cite{sweeney2002k,narayanan2008robust,de2013unique}. Accordingly, PL2 should include not only direct identifiers such as names and contact details, but also stable account identifiers, network identifiers, detailed addresses, and institutional background descriptors that may be weak individually yet highly linkable in combination.

PL3 is defined by expected harm rather than by identifiability alone. Information falls into this tier when its leakage or unlawful use is reasonably expected to cause significant harm to personal safety, property, health, dignity, reputation, or fair opportunity, or when it belongs to categories that law and policy have long treated as specially sensitive. This harm-centered criterion aligns closely with the special categories of personal data under the General Data Protection Regulation, which explicitly single out data such as biometrics, health information, financial accounts, religious belief, sensitive identity information, location, and minors' data because of their elevated risk profile. The broader sensitivity literature likewise emphasizes harm likelihood, confidential relationships, contextual use, and inference risk as central determinants of whether information should be treated as sensitive \cite{belen2022investigation,quinn2021difficulty}. Under this rationale, PL3 encompasses highly consequential records such as government document numbers, financial records, medical data, precise trajectories, biometrics, raw communication content, judicial records, and sensitive attributes, even when some of these items are not uniquely identifying on their own.  

PL4 forms the highest-priority tier and is intentionally stricter than ordinary PII taxonomies \cite{schwartz2011pii,lee2024deepfakes}. Its defining property is immediate exploitability. The exposed artifact is not merely descriptive of a person, but directly reusable for authentication, authorization, signing, privileged access, or system intrusion. This framing is consistent with the digital identity guidance issued by the U.S. National Institute of Standards and Technology, which treats authentication as proof of control over secrets and authenticators, and with operational security guidance that treats secrets as authorization material whose exposure demands immediate revocation and rotation. It is also consistent with threat-intelligence practice. For example, stolen session cookies can be reused to access already authenticated services and may bypass parts of MFA-protected workflows \cite{zheng2015cookies,kondracki2021catching}. For this reason, PL4 includes passwords, PINs, verification codes, session tokens, API keys, private keys, seed phrases, credential-bearing connection strings, exploitable internal access details, and other undisclosed materials whose leakage can directly result in account takeover, financial loss, lateral movement, or large-scale data exfiltration.

\subsection{MemPrivacy-Bench}
MemPrivacy-Bench is introduced to address a central limitation of prior privacy benchmarks for LLM systems: existing resources do not adequately cover the four-level privacy taxonomy required by MemPrivacy, nor do they faithfully reflect the long-term, memory-centric interaction setting of cloud agents. The benchmark is therefore designed for a more realistic problem formulation, where privacy may be disclosed explicitly or implicitly across multi-turn conversations and must be protected without destroying the semantic cues needed for reasoning, retrieval, and personalization. In this sense, MemPrivacy-Bench is not merely a privacy extraction dataset; it is a benchmark for studying the privacy–utility trade-off in cloud agent memory. 

\paragraph{Dataset Construction.}
We synthesize 200 complete user profiles from PersonaHub~\cite{tao2024personahub} seeds. Each profile comprises basic attributes, preferences, and privacy fields spanning all defined levels, with privacy content programmatically sampled to ensure diversity. In particular, each user profile contains preferences in four categories (diet, arts \& entertainment, lifestyle \& hobbies, and aesthetics) and privacy information covering an average of 50 types. Some privacy entries further include multiple fine-grained details, providing rich sources of privacy content. Based on these profiles, we generate multi-turn user--assistant dialogues in which privacy information is revealed both directly and indirectly, simulating realistic conversational exposure. To diversify topics and interaction patterns, we define 7 high-level scenario categories and 23 fine-grained subcategories. The seven high-level categories include drafting \& polishing, financial \& data analysis, consultation \& planning, tech support \& simulation, emotional \& social, contextual inference, and preference change. For each user, we randomly sample 6 to 10 subcategories to generate diverse multi-turn dialogues.

The resulting dataset contains approximately 1M dialogue tokens across balanced Chinese and English data, with each language accounting for 50\% of the dialogues. The training set consists of 26,016 dialogue turns from 160 users, including over 125k privacy instances, while the test set contains 6,337 dialogue turns from 40 users, with over 29.9k privacy instances. Table~\ref{tab:dataset_statistics} summarizes the detailed statistics of the MemPrivacy-Bench training and test sets, as well as the evaluation set derived from PersonaMem-v2~\cite{jiang2025personamemv2}.

\begin{table}[t]
\caption{Statistics of three datasets used in our evaluation.}
    \label{tab:dataset_statistics}
    \centering
    \resizebox{0.6\linewidth}{!}{
    \begin{tabular}{lccc}
    \toprule
    \multirow{2}{*}{\textbf{Metric}} 
    & \multicolumn{2}{c}{\textbf{MemPrivacy-Bench}} 
    & \textbf{PersonaMem-v2} \\
    \cmidrule(lr){2-3} \cmidrule(lr){4-4}
    & \textbf{Train} & \textbf{Test} & \textbf{Evaluation} \\
    \midrule
    \multicolumn{4}{l}{\textit{Overall Scale}} \\
    Number of Users            & 160     & 40      & 20 \\
    Total Turns                & 26,016  & 6,337   & 2,521 \\
    Total Messages             & 52,032  & 12,674  & 5,041 \\
    Messages with Privacy      & 42,159  & 10,234  & 1,085 \\
    Privacy Instances          & 125,776 & 29,967  & 2,378 \\
    Number of Questions        & 0       & 615     & 563 \\
    Total Tokens               & 7,811,374 & 1,872,220 & 671,987 \\
    \midrule
    \multicolumn{4}{l}{\textit{Per-User Averages}} \\
    Turns per User             & 162.60  & 158.43  & 126.03 \\
    Messages per User          & 325.20  & 316.85  & 252.05 \\
    Privacy Instances per User & 786.10  & 749.17  & 118.90 \\
    Questions per User         & 0.00    & 15.38   & 28.15 \\
    Tokens per User            & 48,821.09 & 46,805.50 & 33,599.35 \\
    \midrule
    \multicolumn{4}{l}{\textit{Privacy Level Distribution}} \\
    PL4 Instances              & 19,497  & 4,845   & 22 \\
    PL3 Instances              & 59,724  & 14,562  & 553 \\
    PL2 Instances              & 46,555  & 10,560  & 1,803 \\
    \bottomrule
    \end{tabular}
    }
    \begin{tablenotes}
        \footnotesize
        \item \textit{Note}: - The test split of MemPrivacy-Bench and PersonaMem-v2 include memory question-answer pairs, while the training split does not.
    \end{tablenotes}
\end{table}

\paragraph{Evaluation Construction.}
For downstream evaluation of memory systems, we further derive six types of memory question-answer pairs from the test set, covering basic memory, temporal reasoning, adversarial questioning, dynamic updating, implicit inference, and information aggregation. During question generation, we ensure balanced coverage of users' basic attributes, preferences, and privacy information across different question types. In total, the test set contains 615 question-answer pairs.

\paragraph{LLM-Assisted Privacy Annotation Pipeline.}
We developed a hybrid LLM-assisted annotation pipeline to generate high-quality privacy annotations in a scalable and consistent manner. The pipeline begins with dual-model label generation, where Gemini-3.1-Pro and GPT-5.2 independently produce preliminary privacy annotations for each instance. Their outputs are then integrated through a two-stage refinement process: an initial labeling stage that identifies privacy-relevant content and assigns candidate labels, followed by a correction stage that resolves inconsistencies, revises ambiguous cases, and improves label granularity. This design leverages the complementary strengths of different frontier LLMs while reducing model-specific biases and annotation noise.

To ensure methodological consistency across evaluation resources, both the MemPrivacy-Bench test set and the evaluation split derived from PersonaMem-v2 follow the same annotation strategy. The resulting LLM-generated annotations serve as structured candidates for the subsequent human verification stage, enabling a scalable yet quality-controlled annotation workflow.

\paragraph{Human Annotation Protocol.}
To further ensure the quality and reliability of privacy annotations in the MemPrivacy-Bench test set and the evaluation split derived from PersonaMem-v2, we employed six human annotators to conduct a final round of manual verification and correction after the initial LLM-based labeling produced by Gemini-3.1-Pro and GPT-5.2. All annotators held at least a bachelor's degree and were compensated in accordance with local wage standards. To minimize fatigue-related annotation errors, annotation workloads were scheduled within reasonable weekday working-hour limits.

Before the formal verification stage, all annotators were required to study the privacy taxonomy and extraction guidelines described in Section~\ref{Four-Level Privacy Taxonomy}. This preparation step was intended to align annotation criteria across annotators, improve the consistency of judgments, and reduce subjective variation in borderline cases. The human verification process was performed at the level of individual privacy items. For each annotated item, annotators examined three fields: the \textit{original span text}, the assigned \textit{privacy level}, and the assigned \textit{privacy type}. An annotation was regarded as correct only when all three fields were accurate simultaneously; otherwise, it was marked as incorrect and revised accordingly. This conjunctive criterion makes the verification standard intentionally strict, as partially correct annotations were not accepted.

When ambiguity could not be resolved from the extracted item alone, annotators were allowed to consult the associated user profile information and the original dialogue context. This additional contextual inspection helped ensure that privacy judgments were grounded in the full semantic context rather than in isolated text fragments, which is particularly important for distinguishing fine-grained privacy categories and determining the appropriate privacy level.

After human verification, the privacy annotations in the two evaluation sets achieved an accuracy of 98.08\%. This result provides strong evidence that the overall data construction pipeline is reliable and that the proposed taxonomy is sufficiently well-defined for consistent annotation. More importantly, it suggests that the training split, although not exhaustively manually annotated, remains suitable for training privacy extraction models, while the manually verified evaluation sets provide a dependable basis for benchmark assessment. More details are provided in Appendix~\ref{appendix:detail_of_dataset}.

\subsection{MemPrivacy Model Training}
We train MemPrivacy in two stages, consisting of supervised fine-tuning (SFT) \cite{ouyang2022training} followed by reinforcement learning (RL) with Group Relative Policy Optimization (GRPO) \cite{shao2024deepseekmath}. This two-stage design combines the stability of imitation learning with the flexibility of preference-based policy optimization, enabling the model to acquire strong privacy span extraction capabilities while further improving robustness under fine-grained privacy criteria.

In the first stage, we perform SFT on 26K training instances to establish a reliable initialization for privacy-aware generation. The model is optimized with the standard autoregressive cross-entropy objective:
\[
\mathcal{L}_{\mathrm{SFT}}(\theta) = -\frac{1}{\tau}\sum_{t=1}^{\tau}\log P_\theta\!\left(o_t \mid o_{<t}, s\right),
\]
where $\theta$ denotes the model parameters, $\tau$ is the target sequence length, $s$ is the input sequence, and $o_t$ is the target token at step $t$. This stage equips the model with the basic ability to identify and generate privacy-preserving substitutions, while also providing a stable starting point for subsequent policy optimization. In practice, this initialization is important for reducing instability and output degeneration during early exploration in reinforcement learning.

To further strengthen the model, we conduct a second-stage reinforcement learning phase on an additional 1K training instances using GRPO. Compared with standard PPO-style optimization \cite{schulman2017proximal}, GRPO removes the need for a separately trained value function and instead estimates advantages based on the relative rewards of multiple sampled outputs for the same input. Concretely, for each input, we sample a group of candidate outputs from the old policy, score them with the reward function, normalize the rewards within the group, and optimize the policy using the clipped objective with KL regularization to the reference model:
\begin{equation*}
\begin{aligned}
\mathcal{J}_{\mathrm{RL}}(\theta)
=\ & 
\mathbb{E}\!\left[
q \sim P(Q), \{o_i\}_{i=1}^{G}\sim \pi_{\theta_{\mathrm{old}}}(O\mid q)
\right]
\frac{1}{G}\sum_{i=1}^{G}\frac{1}{|o_i|}
\sum_{t=1}^{|o_i|}\\
&\left\{
\min\!\left[
\frac{\pi_\theta(o_{i,t}\mid q,o_{i,<t})}{\pi_{\theta_{\mathrm{old}}}(o_{i,t}\mid q,o_{i,<t})}\hat{A}_{i,t},
\,
\mathrm{clip}\!\left(
\frac{\pi_\theta(o_{i,t}\mid q,o_{i,<t})}{\pi_{\theta_{\mathrm{old}}}(o_{i,t}\mid q,o_{i,<t})},
1-\epsilon,1+\epsilon
\right)\hat{A}_{i,t}
\right]
-\beta\,\mathbb{D}_{\mathrm{KL}}\!\left[\pi_\theta \,\|\, \pi_{\mathrm{ref}}\right]
\right\}.
\label{eq:grpo_objective}
\end{aligned}
\end{equation*}
where $|o_i|$ is the length of the $i$-th output, $\hat{A}_{i,t}$ denotes the token-level advantage, $\epsilon$ is the PPO-style clipping coefficient, $\pi_{\mathrm{ref}}$ is the reference policy initialized from the SFT model, and $\beta$ controls the strength of the KL regularization term. The ratio
\[
\frac{\pi_\theta(o_{i,t}\mid q,o_{i,<t})}{\pi_{\theta_{\mathrm{old}}}(o_{i,t}\mid q,o_{i,<t})}
\]
measures how the current policy deviates from the behavior policy that generated the sampled outputs, while the KL term constrains the updated policy from drifting excessively away from the SFT initialization.

For reward construction, we use the F1 score defined in our evaluation function (as shown in Section \ref{Experimental Setup}) as the scalar reward for each sampled output. Specifically, given a sampled prediction and its ground-truth annotation, we compute its extraction-level F1 score and treat this value as the reward $r_i$. For each group of $G$ outputs corresponding to the same input, the rewards are normalized by the group mean and standard deviation, yielding
\[
\tilde{r}_i = \frac{r_i - \mathrm{mean}(\mathbf{r})}{\mathrm{std}(\mathbf{r})},
\]
where $\mathbf{r} = \{r_1, \dots, r_G\}$. We then use the normalized reward as the token-level advantage for all tokens in the corresponding output, i.e., $\hat{A}_{i,t}=\tilde{r}_i$. We adopt GRPO because the privacy extraction task benefits from relative comparison among candidate outputs under structured reward signals, while avoiding the computational overhead of critic training. In this way, GRPO directly optimizes the model toward the same criterion used in evaluation, encouraging outputs with higher precision and recall balance in privacy span extraction.

\section{Experiments}
\label{headings}

We evaluate MemPrivacy along the two dimensions defined in Section~\ref{subsec:formal_definition}. To assess privacy leakage risk, we measure the performance of the MemPrivacy model on privacy extraction, namely whether it can accurately identify privacy spans in dialogue together with their privacy levels and privacy types. To assess utility loss, we compare the question-answering performance of memory systems under different privacy protection methods against the no-protection setting. A smaller performance drop indicates better preservation of both system utility and user experience.

\subsection{Experimental Setup}
\label{Experimental Setup}

\paragraph{Datasets.}
We mainly use the test set of \textbf{MemPrivacy-Bench} and an evaluation set constructed from \textbf{PersonaMem-v2}. PersonaMem-v2 is a benchmark for evaluating LLM personalization, with an emphasis on inferring implicit user preferences from long-context conversations. We sample 20 users and their corresponding multiple-choice evaluation questions from the original PersonaMem-v2 dataset; all selected dialogues contain privacy information across different levels. We use MemPrivacy-Bench for in-distribution evaluation and PersonaMem-v2 for out-of-distribution evaluation.

\paragraph{Metric.}
We score privacy extraction using three aspects: \textit{privacy text}, \textit{level}, and \textit{type}. For each matched prediction--reference pair, the final score is the average of the three sub-scores. The original-text score is 1 for an exact match; otherwise, for non-strict types, we use a soft token-level F1 score based on the longest common contiguous substring, and assign 0 to other mismatches (e.g., passwords). Privacy level is scored by exact match. Privacy type is scored by the cosine similarity between type-label embeddings. After greedy one-to-one matching between predictions and references, precision is computed as the average score over predictions, recall as the average score over gold items, and F1 as their harmonic mean. For utility evaluation, question answering on the MemPrivacy-Bench test set is scored by GPT-5.2 as the judge and additionally evaluated with four standard metrics (BLEU-1~\cite{papineni2002bleu}, BLEU-2~\cite{papineni2002bleu}, METEOR~\cite{banerjee2005meteor}, and ROUGE-L~\cite{lin2004rouge}), while the multiple-choice questions from PersonaMem-v2 are evaluated by exact match.

\paragraph{Experimental Configurations}
\label{app:exp_config}
We select LangMem, Mem0, and Memobase as three representative memory systems to study the impact of privacy protection on memory-based agents. Rather than treating them as direct baselines for comparison, we use them as testbeds to examine how privacy protection affects system performance, and whether our method can effectively reduce privacy exposure while preserving utility. For reproducibility, we report the exact code versions, implementation sources, and all non-default configuration changes in the supplementary materials. To ensure fair comparison, all methods use GPT-4.1 for memory operations and question answering whenever applicable. The prompt templates for dataset construction, short-answer evaluation, and privacy information extraction are provided in the supplementary materials. For training, we use LLaMA-Factory \cite{zheng2024llamafactory} for SFT and MS-Swift \cite{zhao2025swift} for RL, with DeepSpeed ZeRO-3 \cite{rajbhandari2020zero} for distributed optimization. In the RL stage, we employ a GRPO-based training objective and use vLLM in a server-based rollout generation setup to improve sampling efficiency. We further incorporate dynamic sampling, bounded resampling, and KL-regularized policy updates to enhance optimization stability. The corresponding training scripts are available in the code repository. All training and evaluation experiments are conducted on a single node equipped with 8 NVIDIA A800-80GB GPUs. Unless otherwise noted, all reported experimental times are measured based on a single NVIDIA A800 GPU.

\subsection{Privacy Extraction Performance}

\begin{table*}[t]
\caption{Performance comparison of different LLMs and privacy models on MemPrivacy-Bench and PersonaMem-v2.}
    \label{tab:model_comparison_memprivacy}
    \centering
    \resizebox{\linewidth}{!}{
    \begin{threeparttable}
        \begin{tabular}{lcccccccc}
        \toprule
        \multicolumn{1}{c}{\multirow{2}{*}{\textbf{Model}}} 
        & \multicolumn{4}{c}{\textbf{MemPrivacy-Bench}} 
        & \multicolumn{4}{c}{\textbf{PersonaMem-v2}} \\
        \cmidrule(lr){2-5} \cmidrule(lr){6-9}
        & \textbf{F1} & \textbf{Precision} & \textbf{Recall} & \textbf{Time (s)}
        & \textbf{F1} & \textbf{Precision} & \textbf{Recall} & \textbf{Time (s)} \\
        \midrule

        \multicolumn{9}{l}{\textbf{General Models}} \\
        \quad Qwen3-0.6B & 26.43 & 21.73 & 41.58 & 3.05 & 9.32 & 8.05 & 13.49 & 2.36 \\
        \quad Qwen3-1.7B & 30.83 & 26.48 & 43.85 & 3.16 & 6.92 & 5.97 & 10.53 & 2.17 \\
        \quad Qwen3-4B & 59.34 & 57.29 & 66.78 & 2.72 & 86.36 & 86.38 & 86.99 & 0.64 \\
        \quad LongCat-Flash-Chat & 64.82 & 65.56 & 67.46 & 4.32 & 86.79 & 87.16 & 86.91 & 1.75 \\
        \quad DeepSeek-V3.1-Terminus & 66.01 & 69.57 & 65.86 & 5.34 & 85.91 & 86.23 & 85.84 & 1.68 \\
        \quad DeepSeek-V3.2 & 65.91 & 68.07 & 67.19 & 7.03 & 86.61 & 86.93 & 86.56 & 3.55 \\
        \quad GPT-5.2 & 68.99 & 65.40 & 78.13 & 4.62 & 88.06 & 87.70 & 89.48 & 3.23 \\
        \quad Kimi-K2 & 69.35 & 73.58 & 69.08 & 5.18 & 87.35 & 88.39 & 86.94 & 1.81 \\
        \quad GLM-5 & 70.10 & 70.13 & 73.29 & 11.30 & 87.87 & 87.97 & 88.27 & 8.56 \\
        \quad DeepSeek-V3.2-Think & 75.04 & 76.46 & 76.13 & 96.14 & 92.18 & 92.16 & 92.84 & 39.60 \\
        \quad Gemini-3.1-Pro & 78.41 & 78.66 & 81.22 & 32.87 & 86.59 & 86.68 & 86.82 & 29.21 \\

        \cdashline{1-9}
        \addlinespace[1pt]
        \multicolumn{9}{l}{\textbf{Privacy Models}} \\
        \quad OpenAI-Privacy-Filter\textsuperscript{*} & 35.50 & 39.96 & 34.30 &  \textbf{0.34} & 85.27 & 85.80 & 85.75 & \textbf{0.24} \\
        \rowcolor[rgb]{0.94,0.97,1.00}
        \quad MemPrivacy-0.6B-SFT & 83.09 & 85.67 & 82.84 & 1.96 & 92.08 & 92.78 & 91.93 & 0.70 \\
        \rowcolor[rgb]{0.94,0.97,1.00}
        \quad MemPrivacy-1.7B-SFT & 84.22 & 86.14 & 84.44 & 2.14 & 93.16 & 93.65 & 93.21 & 0.59 \\
        \rowcolor[rgb]{0.94,0.97,1.00}
        \quad MemPrivacy-4B-SFT & \underline{85.64} & \textbf{87.45} & 86.04 & 2.64 & \underline{94.43} & \underline{94.52} & \underline{95.00} & 0.67 \\
        \rowcolor[rgb]{0.94,0.97,1.00}
        \quad MemPrivacy-0.6B-RL & 84.66 & 86.35 & 85.29 & \underline{1.63} & 93.40 & 94.07 & 93.31 & \underline{0.43} \\
        \rowcolor[rgb]{0.94,0.97,1.00}
        \quad MemPrivacy-1.7B-RL & 84.54 & 85.21 & \underline{86.06} & 1.81 & 94.02 & 94.39 & 94.20 & 0.48 \\
        \rowcolor[rgb]{0.94,0.97,1.00}
        \quad MemPrivacy-4B-RL & \textbf{85.97} & \underline{86.86} & \textbf{87.15} & 2.05 & \textbf{94.48} & \textbf{94.54} & \textbf{95.11} & 0.44 \\
        \bottomrule
        \end{tabular}
        \begin{tablenotes}
            \footnotesize
            \item \textit{Note}: - Best and second-best results are \textbf{bold} and \underline{underlined}, respectively. For inference time, lower is better.
        \end{tablenotes}
    \end{threeparttable}
    }
\end{table*}

Table~\ref{tab:model_comparison_memprivacy} compares 11 general LLMs, one specialized privacy-filtering model, and 6 MemPrivacy models across two datasets. Overall, MemPrivacy consistently outperforms general models by a clear margin on all metrics, demonstrating its strong ability to accurately identify privacy information. The best-performing MemPrivacy model achieves F1 scores of \textbf{85.97\%} and \textbf{94.48\%} on the two datasets, respectively, significantly surpassing the highest F1 scores of general models, which are 78.41\% and 92.18\%. Notably, even the smallest MemPrivacy model (0.6B) attains F1 scores of \textbf{83.09\%} and \textbf{92.08\%}, outperforming nearly all general models. The newly released OpenAI-Privacy-Filter provides an informative comparison with a specialized token-classification baseline. Although it is highly efficient, requiring only 0.34s on MemPrivacy-Bench and 0.24s on PersonaMem-v2, its extraction accuracy is substantially lower. It achieves only 35.50\% F1 on MemPrivacy-Bench, trailing the best MemPrivacy model by 50.47\% and even the smallest MemPrivacy model by 47.59\%. On PersonaMem-v2, it obtains 85.27\% F1, which is also 9.21\% lower than the best MemPrivacy model. This result indicates that efficient token-level filtering alone is insufficient for comprehensive privacy extraction in realistic dialogue scenarios.

\begingroup
\renewcommand{\thefootnote}{\fnsymbol{footnote}}
\footnotetext[1]{OpenAI-Privacy-Filter is a bidirectional token-classification model developed by OpenAI for detecting and masking personally identifiable information in text. Since its open-source release in April 2026, it has supported high-throughput data sanitization workflows and on-premises deployment. The model is available at \url{https://huggingface.co/openai/privacy-filter}.}
\endgroup

The performance gap can be partly attributed to differences in task formulation and privacy taxonomy. OpenAI-Privacy-Filter is designed around a fixed and relatively coarse-grained PII taxonomy, covering only eight privacy categories. Moreover, our evaluation suggests weaker robustness on Chinese conversational data, where privacy expressions are often implicit, context-dependent, and linguistically diverse. General LLMs, relying solely on instructions provided in prompts, also struggle to effectively handle complex privacy information across diverse contexts. In contrast, MemPrivacy is trained on large-scale data to identify fine-grained, multi-level privacy information across heterogeneous conversational scenarios, enabling more comprehensive and robust extraction.

\begin{table}[t]
\caption{Performance comparison of LLM-as-a-Judge evaluations for LLMs and privacy models on MemPrivacy-Bench.}
\label{tab:llm_judge_extract}
\centering
\small
\setlength{\tabcolsep}{8pt}
\renewcommand{\arraystretch}{1.08}

\begin{threeparttable}
\begin{tabular}{lccc}
\toprule
\multicolumn{1}{c}{\multirow{2}{*}{\textbf{Model}}} 
& \multicolumn{3}{c}{\textbf{MemPrivacy-Bench}} \\
\cmidrule(lr){2-4}
& \textbf{GLM-5.1} 
& \textbf{DeepSeek-R1-0528} 
& \textbf{Qwen3-235B-A22B} \\
\midrule
\multicolumn{4}{l}{\textbf{General Models}} \\
\quad GPT-5.2 & 66.54	& 67.94	& 69.30 \\
\quad Kimi-K2 & 63.01	& 65.48	& 67.15  \\
\quad GLM-5 &  65.38	& 66.92	& 68.39  \\
\quad DeepSeek-V3.2-Think &  70.00	& 72.07	& 73.02  \\
\quad Gemini-3.1-Pro &  76.15	& 77.02	& 78.60 \\
\addlinespace[2pt]
\cdashline{1-4}
\addlinespace[1pt]
\multicolumn{4}{l}{\textbf{Privacy Models}} \\
\rowcolor[rgb]{0.94,0.97,1.00}
\quad MemPrivacy-0.6B-SFT &  79.82	& 81.15	& 81.35  \\
\rowcolor[rgb]{0.94,0.97,1.00}
\quad MemPrivacy-1.7B-SFT &  81.02	& 82.57	& 82.68  \\
\rowcolor[rgb]{0.94,0.97,1.00}
\quad MemPrivacy-4B-SFT &  \underline{82.98}	& \underline{84.18}	& \underline{84.18} \\
\rowcolor[rgb]{0.94,0.97,1.00}
\quad MemPrivacy-0.6B-RL &  81.74	& 82.86	& 83.26  \\
\rowcolor[rgb]{0.94,0.97,1.00}
\quad MemPrivacy-1.7B-RL &  82.02	& 82.98	& 83.54  \\
\rowcolor[rgb]{0.94,0.97,1.00}
\quad MemPrivacy-4B-RL &  \textbf{83.85}	& \textbf{84.75}	& \textbf{85.09}  \\
\bottomrule
\end{tabular}
\begin{tablenotes}[flushleft]
\footnotesize
\item \textit{Note}: Best and second-best results are \textbf{bold} and \underline{underlined}, respectively. Scores range from 0 to 1.
\end{tablenotes}
\end{threeparttable}
\end{table}

Substantial gains are also observed when comparing MemPrivacy models with their base models. For instance, after SFT and RL training, Qwen3-4B improves its F1 score on MemPrivacy-Bench from \textbf{59.34\%} to \textbf{85.97\%}, validating the effectiveness of our training framework as well as the high quality ofthe MemPrivacy-Bench training data. Among different MemPrivacy variants, SFT alone already yields strong performance, while incorporating RL further enhances overall performance and generalization ability. For example, MemPrivacy-4B-RL achieves higher F1 scores than MemPrivacy-4B-SFT on both datasets.

\begin{table*}[ht!]
\caption{Performance comparison under different privacy protection methods on three memory systems.}
    \label{tab:privacy_protection_comparison}
    \centering
    \resizebox{\linewidth}{!}{
    \begin{threeparttable}
        \begin{tabular}{llcccccc}
        \toprule
        \multicolumn{1}{c}{\multirow{2}{*}{\textbf{Privacy Protection Method}}} & \multirow{2}{*}{\textbf{Masking Level}} & \multicolumn{5}{c}{\textbf{MemPrivacy-Bench}} & \textbf{PersonaMem-v2} \\
        \cmidrule(lr){3-7} \cmidrule(lr){8-8} 
         & & \textbf{Accuracy} & \textbf{BLEU-1} & \textbf{BLEU-2} & \textbf{METEOR} & \textbf{ROUGE-L} & \textbf{Accuracy} \\
        \midrule

        \multicolumn{8}{c}{\textbf{LangMem}} \\
        \cdashline{1-8}
        \addlinespace[1pt]
        None & -- & \textbf{65.37} {\scriptsize(+0.00)} & \underline{20.92} & \textbf{13.80} & \textbf{20.14} & \textbf{31.50} & \textbf{50.98} {\scriptsize(+0.00)} \\
        \multicolumn{8}{l}{Irreversible Masking} \\
        \quad +MemPrivacy Model & PL2, PL3, PL4 & 38.70 {\scriptsize(-26.67)} & 11.85 & 6.36 & 10.84 & 14.11 & 38.01 {\scriptsize(-12.97)} \\
        \multicolumn{8}{l}{Untyped Placeholder Masking} \\
        \quad +MemPrivacy Model & PL2, PL3, PL4 & 58.70 {\scriptsize(-6.67)} & 18.75 & 11.43 & 15.87 & 25.30 & 48.31 {\scriptsize(-2.67)} \\
        \cdashline{1-8}
        \multicolumn{8}{l}{\cellcolor[rgb]{0.94,0.97,1.00}MemPrivacy} \\
        \quad \cellcolor[rgb]{0.94,0.97,1.00}+DeepSeek-V3.2-Think & \cellcolor[rgb]{0.94,0.97,1.00}PL2, PL3, PL4 & \cellcolor[rgb]{0.94,0.97,1.00}58.05 {\scriptsize(-7.32)} & \cellcolor[rgb]{0.94,0.97,1.00}15.88 & \cellcolor[rgb]{0.94,0.97,1.00}9.41 & \cellcolor[rgb]{0.94,0.97,1.00}13.93 & \cellcolor[rgb]{0.94,0.97,1.00}25.18 & \cellcolor[rgb]{0.94,0.97,1.00}46.76 {\scriptsize(-4.22)} \\
        \quad \cellcolor[rgb]{0.94,0.97,1.00}+GPT-5.2 & \cellcolor[rgb]{0.94,0.97,1.00}PL2, PL3, PL4 & \cellcolor[rgb]{0.94,0.97,1.00}54.03 {\scriptsize(-11.34)} & \cellcolor[rgb]{0.94,0.97,1.00}14.91 & \cellcolor[rgb]{0.94,0.97,1.00}7.65 & \cellcolor[rgb]{0.94,0.97,1.00}13.18 & \cellcolor[rgb]{0.94,0.97,1.00}22.75 & \cellcolor[rgb]{0.94,0.97,1.00}47.48 {\scriptsize(-3.50)} \\
        \quad \cellcolor[rgb]{0.94,0.97,1.00}+MemPrivacy Model 
        & \cellcolor[rgb]{0.94,0.97,1.00}PL2, PL3, PL4 
        & \cellcolor[rgb]{0.94,0.97,1.00}64.07 {\scriptsize(-1.30)} 
        & \cellcolor[rgb]{0.94,0.97,1.00}18.64 
        & \cellcolor[rgb]{0.94,0.97,1.00}12.07 
        & \cellcolor[rgb]{0.94,0.97,1.00}17.95 
        & \cellcolor[rgb]{0.94,0.97,1.00}27.91 
        & \cellcolor[rgb]{0.94,0.97,1.00}49.38 {\scriptsize(-1.60)} \\
        \quad \cellcolor[rgb]{0.94,0.97,1.00}+MemPrivacy Model 
        & \cellcolor[rgb]{0.94,0.97,1.00}PL3, PL4 
        & \cellcolor[rgb]{0.94,0.97,1.00}65.12 {\scriptsize(-0.25)} 
        & \cellcolor[rgb]{0.94,0.97,1.00}19.47 
        & \cellcolor[rgb]{0.94,0.97,1.00}12.72 
        & \cellcolor[rgb]{0.94,0.97,1.00}\underline{18.27} 
        & \cellcolor[rgb]{0.94,0.97,1.00}\underline{29.83} 
        & \cellcolor[rgb]{0.94,0.97,1.00}49.91 {\scriptsize(-1.07)} \\
        \quad \cellcolor[rgb]{0.94,0.97,1.00}+MemPrivacy Model 
        & \cellcolor[rgb]{0.94,0.97,1.00}PL4 
        & \cellcolor[rgb]{0.94,0.97,1.00}\underline{65.28} {\scriptsize(-0.09)} 
        & \cellcolor[rgb]{0.94,0.97,1.00}\textbf{22.25} 
        & \cellcolor[rgb]{0.94,0.97,1.00}\underline{13.09} 
        & \cellcolor[rgb]{0.94,0.97,1.00}18.03 
        & \cellcolor[rgb]{0.94,0.97,1.00}29.35 
        & \cellcolor[rgb]{0.94,0.97,1.00}\underline{50.80} {\scriptsize(-0.18)} \\

        \midrule

        \multicolumn{8}{c}{\textbf{Mem0}} \\
        \cdashline{1-8}
        \addlinespace[1pt]
        None & -- & \textbf{68.62} {\scriptsize(+0.00)} & \textbf{15.40} & \textbf{10.05} & \textbf{15.60} & \textbf{25.42} & \textbf{55.42} {\scriptsize(+0.00)} \\
        \multicolumn{8}{l}{Irreversible Masking} \\
        \quad +MemPrivacy Model & PL2, PL3, PL4 & 26.75 {\scriptsize(-41.87)} & 11.07 & 6.36 & 12.34 & 13.85 & 44.05 {\scriptsize(-11.37)} \\
        Untyped Placeholder Masking & \multicolumn{7}{l}{} \\
        \quad +MemPrivacy Model & PL2, PL3, PL4 & 63.90 {\scriptsize(-4.72)} & 12.34 & 7.81 & 14.09 & 22.56 & 46.71 {\scriptsize(-8.71)} \\
        \cdashline{1-8}
        \multicolumn{8}{l}{\cellcolor[rgb]{0.94,0.97,1.00}MemPrivacy} \\
        \quad \cellcolor[rgb]{0.94,0.97,1.00}+DeepSeek-V3.2-Think & \cellcolor[rgb]{0.94,0.97,1.00}PL2, PL3, PL4 & \cellcolor[rgb]{0.94,0.97,1.00}37.58 {\scriptsize(-31.04)} & \cellcolor[rgb]{0.94,0.97,1.00}9.32 & \cellcolor[rgb]{0.94,0.97,1.00}5.22 & \cellcolor[rgb]{0.94,0.97,1.00}9.28 & \cellcolor[rgb]{0.94,0.97,1.00}18.46 & \cellcolor[rgb]{0.94,0.97,1.00}49.28 {\scriptsize(-6.14)} \\
        \quad \cellcolor[rgb]{0.94,0.97,1.00}+GPT-5.2 & \cellcolor[rgb]{0.94,0.97,1.00}PL2, PL3, PL4 & \cellcolor[rgb]{0.94,0.97,1.00}32.21 {\scriptsize(-36.41)} & \cellcolor[rgb]{0.94,0.97,1.00}6.93 & \cellcolor[rgb]{0.94,0.97,1.00}3.57 & \cellcolor[rgb]{0.94,0.97,1.00}6.99 & \cellcolor[rgb]{0.94,0.97,1.00}16.56 & \cellcolor[rgb]{0.94,0.97,1.00}50.36 {\scriptsize(-5.06)} \\
        \quad \cellcolor[rgb]{0.94,0.97,1.00}+MemPrivacy Model 
        & \cellcolor[rgb]{0.94,0.97,1.00}PL2, PL3, PL4 
        & \cellcolor[rgb]{0.94,0.97,1.00}67.89 {\scriptsize(-0.73)} 
        & \cellcolor[rgb]{0.94,0.97,1.00}14.59 
        & \cellcolor[rgb]{0.94,0.97,1.00}9.04 
        & \cellcolor[rgb]{0.94,0.97,1.00}14.97 
        & \cellcolor[rgb]{0.94,0.97,1.00}23.78 
        & \cellcolor[rgb]{0.94,0.97,1.00}54.71 {\scriptsize(-0.71)} \\
        \quad \cellcolor[rgb]{0.94,0.97,1.00}+MemPrivacy Model 
        & \cellcolor[rgb]{0.94,0.97,1.00}PL3, PL4 
        & \cellcolor[rgb]{0.94,0.97,1.00}67.97 {\scriptsize(-0.65)} 
        & \cellcolor[rgb]{0.94,0.97,1.00}13.75 
        & \cellcolor[rgb]{0.94,0.97,1.00}8.95 
        & \cellcolor[rgb]{0.94,0.97,1.00}14.54 
        & \cellcolor[rgb]{0.94,0.97,1.00}23.78 
        & \cellcolor[rgb]{0.94,0.97,1.00}\underline{54.88} {\scriptsize(-0.54)} \\
        \quad \cellcolor[rgb]{0.94,0.97,1.00}+MemPrivacy Model 
        & \cellcolor[rgb]{0.94,0.97,1.00}PL4 
        & \cellcolor[rgb]{0.94,0.97,1.00}\underline{68.29} {\scriptsize(-0.33)} 
        & \cellcolor[rgb]{0.94,0.97,1.00}\underline{15.34} 
        & \cellcolor[rgb]{0.94,0.97,1.00}\underline{9.51} 
        & \cellcolor[rgb]{0.94,0.97,1.00}\underline{15.39} 
        & \cellcolor[rgb]{0.94,0.97,1.00}\underline{24.72} 
        & \cellcolor[rgb]{0.94,0.97,1.00}\textbf{55.42} {\scriptsize(+0.00)} \\

        \midrule

        \multicolumn{8}{c}{\textbf{Memobase}} \\
        \cdashline{1-8}
        \addlinespace[1pt]
        None & -- & \textbf{38.62} {\scriptsize(+0.00)} & \underline{14.76} & \textbf{8.86} & \underline{13.22} & \textbf{22.56} & \textbf{54.35} {\scriptsize(+0.00)} \\
        \multicolumn{8}{l}{Irreversible Masking} \\
        \quad +MemPrivacy Model & PL2, PL3, PL4 & 21.63 {\scriptsize(-16.99)} & 9.21 & 4.96 & 9.84 & 12.14 & 46.18 {\scriptsize(-8.17)} \\
        \multicolumn{8}{l}{Untyped Placeholder Masking} \\
        \quad +MemPrivacy Model & PL2, PL3, PL4 & 33.90 {\scriptsize(-4.72)} & 11.27 & 6.61 & 11.39 & 18.72 & 48.49 {\scriptsize(-5.86)} \\
        \cdashline{1-8}
        \multicolumn{8}{l}{\cellcolor[rgb]{0.94,0.97,1.00}MemPrivacy} \\
        \quad \cellcolor[rgb]{0.94,0.97,1.00}+DeepSeek-V3.2-Think & \cellcolor[rgb]{0.94,0.97,1.00}PL2, PL3, PL4 & \cellcolor[rgb]{0.94,0.97,1.00}25.50 {\scriptsize(-13.12)} & \cellcolor[rgb]{0.94,0.97,1.00}6.60 & \cellcolor[rgb]{0.94,0.97,1.00}3.30 & \cellcolor[rgb]{0.94,0.97,1.00}5.61 & \cellcolor[rgb]{0.94,0.97,1.00}11.25 & \cellcolor[rgb]{0.94,0.97,1.00}47.12 {\scriptsize(-7.23)} \\
        \quad \cellcolor[rgb]{0.94,0.97,1.00}+GPT-5.2 & \cellcolor[rgb]{0.94,0.97,1.00}PL2, PL3, PL4 & \cellcolor[rgb]{0.94,0.97,1.00}33.56 {\scriptsize(-5.06)} & \cellcolor[rgb]{0.94,0.97,1.00}5.87 & \cellcolor[rgb]{0.94,0.97,1.00}2.77 & \cellcolor[rgb]{0.94,0.97,1.00}5.97 & \cellcolor[rgb]{0.94,0.97,1.00}11.19 & \cellcolor[rgb]{0.94,0.97,1.00}46.40 {\scriptsize(-7.95)} \\
        \quad \cellcolor[rgb]{0.94,0.97,1.00}+MemPrivacy Model 
        & \cellcolor[rgb]{0.94,0.97,1.00}PL2, PL3, PL4 
        & \cellcolor[rgb]{0.94,0.97,1.00}37.89 {\scriptsize(-0.73)} 
        & \cellcolor[rgb]{0.94,0.97,1.00}12.28 
        & \cellcolor[rgb]{0.94,0.97,1.00}7.52 
        & \cellcolor[rgb]{0.94,0.97,1.00}12.11 
        & \cellcolor[rgb]{0.94,0.97,1.00}19.63 
        & \cellcolor[rgb]{0.94,0.97,1.00}53.11 {\scriptsize(-1.24)} \\
        \quad \cellcolor[rgb]{0.94,0.97,1.00}+MemPrivacy Model 
        & \cellcolor[rgb]{0.94,0.97,1.00}PL3, PL4 
        & \cellcolor[rgb]{0.94,0.97,1.00}38.13 {\scriptsize(-0.49)} 
        & \cellcolor[rgb]{0.94,0.97,1.00}12.80 
        & \cellcolor[rgb]{0.94,0.97,1.00}7.50 
        & \cellcolor[rgb]{0.94,0.97,1.00}12.56 
        & \cellcolor[rgb]{0.94,0.97,1.00}20.88 
        & \cellcolor[rgb]{0.94,0.97,1.00}53.29 {\scriptsize(-1.06)} \\
        \quad \cellcolor[rgb]{0.94,0.97,1.00}+MemPrivacy Model 
        & \cellcolor[rgb]{0.94,0.97,1.00}PL4 
        & \cellcolor[rgb]{0.94,0.97,1.00}\underline{38.54} {\scriptsize(-0.08)} 
        & \cellcolor[rgb]{0.94,0.97,1.00}\textbf{14.94} 
        & \cellcolor[rgb]{0.94,0.97,1.00}\underline{8.72} 
        & \cellcolor[rgb]{0.94,0.97,1.00}\textbf{13.54} 
        & \cellcolor[rgb]{0.94,0.97,1.00}\underline{21.44} 
        & \cellcolor[rgb]{0.94,0.97,1.00}\underline{53.46} {\scriptsize(-0.89)} \\

        \bottomrule
        \end{tabular}
        \begin{tablenotes}
            \footnotesize
            \item \textit{Note}: - Numbers in parentheses denote the difference between each method and the corresponding "None" (no privacy protection strategy) baseline under the same memory system. Best and second-best results are \textbf{bold} and \underline{underlined}, respectively.
        \end{tablenotes}
    \end{threeparttable}
    }
\end{table*}

We also observe that all models perform worse on MemPrivacy-Bench than on PersonaMem-v2. This is expected, as MemPrivacy-Bench is intentionally designed as a more challenging stress test, with substantially higher privacy density in dialogues, containing over 29.9k privacy instances in the test set, compared to only 2.3k+ in PersonaMem-v2. By contrast, the privacy density in PersonaMem-v2 is closer to that of real-world user–assistant conversations, and therefore better reflects practical deployment conditions. In addition to accuracy gains, MemPrivacy models are also significantly more efficient. On PersonaMem-v2, the latency for processing a single message is consistently \textbf{below one second}, and even on the privacy-dense MemPrivacy-Bench, it remains around two seconds. In comparison, large reasoning models such as Gemini-3.1-Pro are one to two orders of magnitude slower. This latency profile makes MemPrivacy well suited for on-device deployment without introducing noticeable interaction delays.

Beyond direct benchmark performance and inference efficiency, we further examine whether MemPrivacy models can provide reliable privacy judgments when evaluated by external strong LLM judges. Table~\ref{tab:llm_judge_extract} reports an LLM-as-a-Judge evaluation on MemPrivacy-Bench, where different judge backbones are used to assess the quality of privacy-related decisions. The results show a clear and consistent advantage of privacy-specialized models over general-purpose LLMs across all three judge models. Even the smallest MemPrivacy model, MemPrivacy-0.6B-SFT, surpasses the strongest general baseline across all evaluators, indicating that task-specific privacy alignment provides substantial benefits beyond model scale alone. Performance improves steadily as model size increases from 0.6B to 4B, and reinforcement learning further enhances the models compared with their SFT counterparts. Notably, MemPrivacy-4B-RL achieves the best results across all judge models, reaching 83.85\%, 84.75\%, and 85.09\%, respectively, while MemPrivacy-4B-SFT obtains the second-best performance. These findings suggest that the proposed training strategy enables smaller privacy-oriented models to develop more reliable evaluation capabilities than much larger general-purpose LLMs, highlighting the importance of domain-specific supervision and reinforcement learning for privacy-aware evaluation.

\subsection{Memory System Performance under Different Privacy Protection methods} 

To evaluate the practical impact of privacy protection on cloud agents memory, we test three widely used memory systems, all backed by GPT-4.1 with consistent retrieval and answering settings. We compare MemPrivacy against two representative baselines. \textbf{Irreversible masking} replaces all privacy spans with mask token \textbf{***} and performs no response-side restoration, while \textbf{untyped placeholder masking} replaces each privacy span with a generic placeholder such as \texttt{<Mask\_1>}, without preserving semantic type information. For both baselines, privacy spans are extracted by the MemPrivacy model.

Table~\ref{tab:privacy_protection_comparison} shows that MemPrivacy preserves memory-system utility much better than the baselines. When all PL2--PL4 privacy content is protected, the performance drop under MemPrivacy remains very small across all systems, only \textbf{0.71\%--1.60\%} on MemPrivacy-Bench and PersonaMem-v2. As the masking level becomes more selective, the loss further decreases. When only PL4 is protected, the drop is below \textbf{0.89\%} on MemPrivacy-Bench and PersonaMem-v2. Beyond accuracy, the generation-quality metrics exhibit the same trend. Across BLEU-1, BLEU-2, METEOR, and ROUGE-L, MemPrivacy consistently maintains higher response fidelity than irreversible masking and untyped placeholder masking, indicating that type-aware reversible protection better preserves the lexical and semantic information needed for memory retrieval and answer generation. In contrast, irreversible masking causes the largest degradation on all systems, since it removes both private values and semantic roles. Untyped placeholder masking performs better than irreversible masking but still lags far behind MemPrivacy, showing that generic placeholders preserve too little information for reliable memory retrieval and response generation. 

We also compare different privacy extractors within the MemPrivacy framework. Replacing the MemPrivacy model with strong general models such as DeepSeek-V3.2-Think or GPT-5.2 still leads to substantial utility loss. On Mem0 over MemPrivacy-Bench, their accuracies drop to \textbf{37.58\%} and \textbf{32.21\%}, far below the \textbf{67.89\%} achieved by MemPrivacy with its own model. The same pattern is also reflected in generation quality: for example, on Mem0, the ROUGE-L score decreases from \textbf{23.78\%} with the MemPrivacy extractor to \textbf{18.46\%} and \textbf{16.56\%} with DeepSeek-V3.2-Think and GPT-5.2, respectively. This indicates that the effectiveness of the overall framework critically depends on accurate privacy extraction. Overall, these results show that MemPrivacy achieves a much better privacy--utility trade-off than simpler masking strategies and remains directly applicable to real memory-based agent systems.

\begin{figure}[t]
    \centering
    \includegraphics[width=\linewidth]{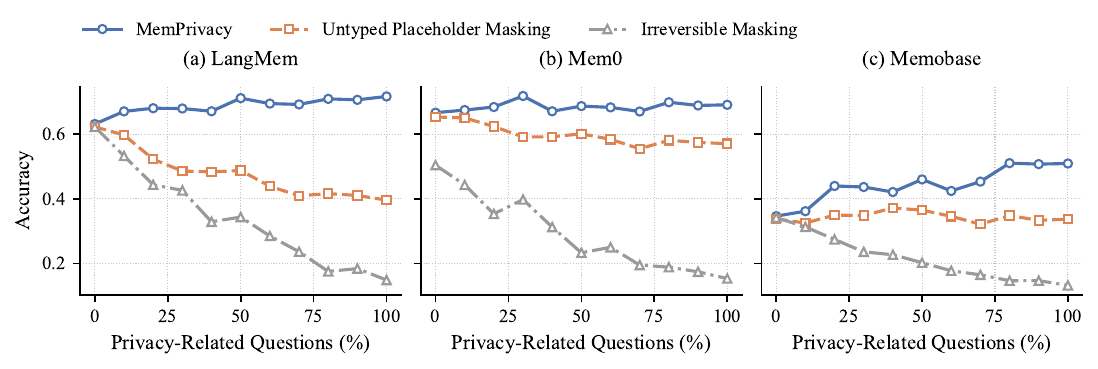}
    \caption{Overall accuracy on three memory systems under different proportions of privacy-related questions.}
    \label{fig:privacy_related_acc}
\end{figure}

Figure~\ref{fig:privacy_related_acc} further evaluates robustness under different proportions of privacy-related questions, where 0\% denotes purely non-privacy questions and 100\% denotes fully privacy-related questions. Across all three memory systems, MemPrivacy remains consistently more stable as the proportion of privacy-related questions increases, while the gap between MemPrivacy and the baselines becomes progressively larger in privacy-heavy settings.

This trend reflects the core advantage of MemPrivacy. By preserving semantic type information through typed placeholders, it allows memory writing, retrieval, and answer generation to remain effective even when more questions depend on privacy-bearing content. In contrast, untyped placeholder masking provides only limited semantic cues and therefore shows a clearer performance decline, whereas irreversible masking causes the most severe degradation because it removes both the private values and their semantic roles. Overall, these results further confirm that MemPrivacy achieves a substantially better privacy--utility trade-off, especially in privacy-intensive scenarios where robust protection is most needed.

\section{Conclusion}
We present \textbf{MemPrivacy}, a privacy-preserving personalized memory management framework for edge-cloud agents. MemPrivacy performs local privacy span extraction and replaces sensitive content with semantically meaningful typed placeholders, enabling cloud-side reasoning and memory operations without exposing raw private values, and restores original content locally to preserve user experience. To make this framework practical, we develop a family of privacy extraction models tailored to the memory setting, and further improved the strongest variant through reinforcement learning. Beyond the framework itself, we formalize a four-level privacy taxonomy that captures varying sensitivity and exploitability, and built \textbf{MemPrivacy-Bench}, a dedicated benchmark for privacy-aware cloud memory, covering 200 users and more than 155k+ privacy instances. Extensive experiments across multiple memory systems demonstrate that MemPrivacy consistently outperforms irreversible redaction and untyped placeholder baselines in both privacy extraction quality and downstream memory utility. These results show that MemPrivacy offers a practical path toward privacy-preserving long-term memory for cloud agents.

\bibliographystyle{plainnat}
\bibliography{main}

\clearpage

\newpage
\appendix
\appendixpage

\startcontents[sections]
\printcontents[sections]{l}{1}{\setcounter{tocdepth}{2}}

\section{Supplementary Details of Dataset}
\label{appendix:detail_of_dataset}

This appendix provides detailed statistics of the datasets used in our experiments, together with additional details on the construction of \textbf{MemPrivacy-Bench}. Table~\ref{tab:dataset_statistics} reports the detailed statistics of the \textbf{MemPrivacy-Bench} training and test sets, as well as the evaluation set constructed from \textbf{PersonaMem-v2}.

The training and test splits of MemPrivacy-Bench are balanced across Chinese and English, with each language accounting for 50\% of the dialogues. Each complete user profile includes basic attributes, preferences in four categories (diet, arts and entertainment, lifestyle and hobbies, and aesthetics), and privacy information covering an average of 50 types. Some privacy entries further contain multiple fine-grained details, providing rich sources of privacy content for dialogue generation.

To diversify topics and interaction settings, we define 7 high-level scenario categories and 23 fine-grained subcategories of user--assistant conversations, aiming to better cover realistic situations in which privacy may be revealed. The seven high-level categories are \textit{Drafting \& Polishing}, \textit{Financial \& Data Analysis}, \textit{Consultation \& Planning}, \textit{Tech Support \& Simulation}, \textit{Emotional \& Social}, \textit{Contextual Inference}, and \textit{Preference Change}. For each user, we randomly sample 6 to 10 subcategories and generate the corresponding multi-turn dialogues.

For the test split, we further construct memory question-answer pairs in six common categories: \textit{Basic Memory}, \textit{Temporal Reasoning}, \textit{Adversarial Questioning}, \textit{Dynamic Updating}, \textit{Implicit Inference}, and \textit{Information Aggregation}. During question generation, we ensure coverage of users' basic attributes, preferences, and privacy information, so that both question types and content remain balanced.

Additionally, it should be noted that the privacy information in both the MemPrivacy-Bench test set and the PersonaMem-v2 evaluation set is annotated using the same strategy, with initial labeling performed by Gemini-3.1-Pro and GPT-5.2 and final annotation and verification conducted by human annotators.

\section{Human Annotation Protocol}
\label{app:human_annotation}

To further ensure the annotation quality of privacy information in the MemPrivacy-Bench test set and the evaluation split constructed from PersonaMem-v2, we recruit six human annotators to conduct final verification and correction after the initial LLM-based labeling by Gemini-3.1-Pro and GPT-5.2. All annotators hold at least a bachelor's degree and are compensated according to local wage standards, with workloads scheduled within reasonable working-hour limits on weekdays.

Before annotation, all annotators study the privacy taxonomy and extraction guidelines defined in Sec.~2.1 to align their decision criteria and reduce subjective inconsistency. During verification, annotators review every annotated privacy item in the dialogues, including the \textit{original span text}, \textit{privacy level}, and \textit{privacy type}. An item is considered correct only if all three fields are accurate; otherwise, it is marked as incorrect and revised. When necessary, we provide annotators with the associated user profile information and the original dialogue context to support consistent judgments.

After human verification, the privacy annotations in the two evaluation sets reach an accuracy of 98.08\%. This result indicates that our construction pipeline and taxonomy are reliable, and that the training split can effectively support privacy extraction model training even without exhaustive human annotation.

\section{MemPrivacy Model Variants and Training Recipes}
\label{app:memprivacy_variants}

Table~\ref{tab:memprivacy_mapping} summarizes all MemPrivacy model variants used in our experiments, including their corresponding base models and training recipes. We adopt Qwen3 models at different scales as backbones and train MemPrivacy models with supervised fine-tuning (SFT). For the instruction-tuned 4B backbone, we additionally report a variant trained with SFT followed by reinforcement learning (SFT+RL). Unless otherwise specified, results reported in the main paper use the MemPrivacy model variant indicated in Table~\ref{tab:memprivacy_mapping}.

Model names follow the convention \texttt{MemPrivacy-\{size\}-\{recipe\}}, where \texttt{Ins} denotes instruction-tuned backbones.

\begin{table}[t]
\caption{Mapping between MemPrivacy models, their base models, and training methods.}
\label{tab:memprivacy_mapping}
\centering
\small
\resizebox{0.6\linewidth}{!}{%
\begin{tabular}{llc}
\toprule
\textbf{MemPrivacy Model} & \textbf{Base Model} & \textbf{Training Method} \\
\midrule
MemPrivacy-0.6B-SFT & Qwen3-0.6B & SFT \\
MemPrivacy-0.6B-RL & Qwen3-0.6B & SFT + RL \\
MemPrivacy-1.7B-SFT & Qwen3-1.7B & SFT \\
MemPrivacy-1.7B-RL & Qwen3-1.7B & SFT + RL \\
MemPrivacy-4B-SFT & Qwen3-4B & SFT \\
MemPrivacy-4B-RL & Qwen3-4B & SFT + RL \\
\bottomrule
\end{tabular}
}
\end{table}

\section{Detailed Experimental Configurations}
\label{app:exp_config}
\subsection{Prompt Engineering Details}
To ensure reproducibility and facilitate closer inspection of our evaluation protocol, we provide the prompt templates used at different stages of the benchmark pipeline. The prompts for answering short-answer questions in MemPrivacy-Bench and multiple-choice questions in PersonaMem-v2 are shown in Figures~\ref{fig:prompt_for_qa}-\ref{fig:prompt_for_mcq}. The prompt template used to instruct GPT-5.2 to evaluate memory-system performance on short-answer questions is presented in Figures~\ref{fig:prompt_for_judge_1}-\ref{fig:prompt_for_judge_2}. The prompt template used for privacy information extraction is provided in Figures~\ref{fig:prompt_privacy_extraction_1}-\ref{fig:prompt_privacy_extraction_4}.

In addition, we provide tables containing illustrative examples of detailed privacy policy provisions (Table~\ref{tab:detailed_privacy_policy}), as well as hierarchical privacy level definitions and their associated default handling policies (Table~\ref{tab:privacy_governance_principles}). Together, these materials offer a more complete account of the prompting setup, policy grounding, and annotation principles that support our benchmark construction and evaluation.

\subsection{Response Generation Quality Metrics.}
In addition to task-level accuracy, we report four automatic generation-quality metrics, namely BLEU-1, BLEU-2, METEOR, and ROUGE-L, to evaluate the lexical and sequence-level consistency between generated answers and reference answers. For each question-answer instance, let $\hat{y}$ denote the generated answer and $y$ denote the corresponding reference answer. All metrics are computed between $\hat{y}$ and $y$ using the same tokenization procedure across all evaluated methods, and the final score is obtained by averaging over all test instances. Higher values indicate better agreement with the reference answer.

\textbf{BLEU-1 and BLEU-2.}
BLEU measures modified $n$-gram precision between a candidate answer and its reference, together with a brevity penalty to discourage overly short generations. For BLEU-$N$, the score is defined as
\begin{equation}
\mathrm{BLEU}\text{-}N=\mathrm{BP}\cdot\exp\left(
\sum_{n=1}^{N} w_n \log p_n\right),
\end{equation}
where $p_n$ is the modified precision of $n$-grams, and $w_n$ is the weight assigned to the $n$-gram order. In our evaluation, BLEU-1 uses $N=1$ with $w_1=1$, while BLEU-2 uses $N=2$ with uniform weights $w_1=w_2=\frac{1}{2}$. The brevity penalty is computed as
\begin{equation}
\mathrm{BP}=
\begin{cases}
1, & c > r, \\
\exp(1-r/c), & c \leq r,
\end{cases}
\end{equation}
where $c$ and $r$ denote the lengths of the generated answer and the reference answer, respectively. The modified $n$-gram precision is given by
\begin{equation}
p_n=\frac{
\sum_{g \in \mathcal{G}_n(\hat{y})}
\min \left(
\mathrm{Count}_{\hat{y}}(g),
\mathrm{Count}_{y}(g)
\right)
}{
\sum_{g \in \mathcal{G}_n(\hat{y})}
\mathrm{Count}_{\hat{y}}(g)
},
\end{equation}
where $\mathcal{G}_n(\hat{y})$ denotes the set of $n$-grams in the generated answer, and $\mathrm{Count}_{\hat{y}}(g)$ and $\mathrm{Count}_{y}(g)$ denote the occurrence counts of $n$-gram $g$ in the generated and reference answers, respectively. BLEU-1 mainly captures unigram-level lexical overlap, whereas BLEU-2 additionally evaluates local word-order consistency through bigram overlap.

\textbf{METEOR.}
METEOR evaluates answer quality based on unigram alignment between the generated and reference answers. Compared with pure precision-oriented metrics, METEOR incorporates both precision and recall and penalizes fragmented alignments. Let $m$ be the number of matched unigrams between $\hat{y}$ and $y$. The unigram precision and recall are defined as
\begin{equation}
P = \frac{m}{|\hat{y}|},
\qquad
R = \frac{m}{|y|}.
\end{equation}
The harmonic mean used by METEOR is computed as
\begin{equation}
F_{\mathrm{mean}}=
\frac{10PR}{R + 9P}.
\end{equation}
To account for word-order fragmentation, METEOR applies a penalty term:
\begin{equation}
\mathrm{Penalty}=
0.5\left(\frac{ch}{m}\right)^3,
\end{equation}
where $ch$ denotes the number of contiguous matched chunks. The final METEOR score is
\begin{equation}
\mathrm{METEOR}=F_{\mathrm{mean}}\cdot\left(1-\mathrm{Penalty}\right).
\end{equation}
When no unigram match exists, the METEOR score is set to zero. This metric is useful for evaluating whether the generated answer preserves the key content of the reference answer while allowing moderate variation in surface realization.

\textbf{ROUGE-L.}
ROUGE-L evaluates the longest common subsequence between the generated answer and the reference answer. Unlike fixed-order $n$-gram metrics, ROUGE-L captures in-sequence overlap without requiring consecutive token matches. Let $\mathrm{LCS}(\hat{y}, y)$ denote the length of the longest common subsequence between $\hat{y}$ and $y$. The ROUGE-L recall and precision are defined as
\begin{equation}
R_{\mathrm{LCS}}=\frac{\mathrm{LCS}(\hat{y}, y)}{|y|},\qquad P_{\mathrm{LCS}}
=\frac{\mathrm{LCS}(\hat{y}, y)}{|\hat{y}|}.
\end{equation}
The final ROUGE-L score is computed as the $F$-measure:
\begin{equation}
\mathrm{ROUGE}\text{-}\mathrm{L}=
\frac{
(1+\beta^2) R_{\mathrm{LCS}} P_{\mathrm{LCS}}
}{
R_{\mathrm{LCS}} + \beta^2 P_{\mathrm{LCS}}
}.
\end{equation}
Following the common F1-based setting, we set $\beta=1$ in our experiments. ROUGE-L therefore measures the extent to which the generated answer preserves the global token sequence of the reference answer.

All reported BLEU-1, BLEU-2, METEOR, and ROUGE-L scores are averaged over the evaluation set. In the result tables, we report these scores as percentages for readability.

\subsection{Memory Agent Testbeds and Deployment Configurations}
To systematically examine the impact of privacy-preserving mechanisms on memory-based agents, this paper selects \textbf{LangMem, Mem0, and Memobase} as three representative long-term memory systems. Rather than treating them as homogeneous baselines for direct comparison, we regard them as exemplifying distinct paradigms of memory modeling and system integration: LangMem emphasizes deep coupling with agent frameworks and a typology of memory; Mem0 focuses on a general-purpose memory layer independent of upper-level agents and an extensible retrieval pipeline; and Memobase highlights user-profile-centered long-term personalization with time-aware memory. Precisely because these systems differ substantially in memory representation, writing mechanisms, retrieval strategies, and deployment forms, they provide complementary experimental testbeds for studying the trade-off between privacy exposure and task utility.

\textbf{LangMem.} The core objective of LangMem is to enable agents to continuously learn and adapt through interaction. It not only supports extracting salient information from conversations to form long-term memory, but also supports optimizing the system prompt based on interaction trajectories, thereby unifying \emph{what to remember} and \emph{how to improve behavior accordingly} within a single framework. At the level of memory modeling, LangMem explicitly distinguishes among semantic memory, episodic memory, and procedural memory. Semantic memory stores user preferences, facts, and background knowledge, and may be represented either as a retrievable collection or as a structured profile. Episodic memory preserves successful past interaction cases and the circumstances under which they succeeded. Procedural memory, in turn, treats system instructions as behavioral rules that can be continuously refined through feedback and experience. This typological design allows LangMem to jointly cover three levels of functionality: fact retention, experience transfer, and behavioral calibration. It is therefore particularly relevant to privacy research, since private information may be exposed not only as factual entries, but also indirectly encoded through successful cases, behavioral rules, or even prompt updates.

With respect to memory writing, LangMem supports both \emph{hot-path} and \emph{background reflection} mechanisms for memory formation. The former allows the agent to actively invoke memory tools during an ongoing dialogue to create, update, or delete memories. The latter enables the system, after the conversation ends or during idle periods, to automatically extract, consolidate, and update memory through \texttt{create\_memory\_store\_manager} and a reflection execution mechanism. This design implies that privacy-preserving methods must address not only the exposure risk of explicit online writes, but also the secondary exposure risk introduced by offline reflective distillation.

\textbf{Mem0.} Mem0 defines itself as a universal, self-improving memory layer for LLM applications. Its design goal is to provide persistent context across multi-turn and multi-session settings without requiring developers to bind long-term memory capabilities to any specific agent framework. Its core workflow centers on memory writing, retrieval, and reuse. The search interface supports natural-language queries, JSON-logic-based filtering, \texttt{top\_k} and threshold control, as well as optional reranking for secondary reordering of retrieved results. More importantly, the current implementation of Mem0 allows new facts and old facts to coexist, without proactively overwriting or deleting existing memories, in order to preserve temporal context and avoid information loss caused by premature consolidation. For privacy research, this strategy has a dual implication: on the one hand, it helps retain temporal evidence; on the other hand, it may increase the likelihood that stale yet sensitive information remains in the system for extended periods and is later recalled.

\textbf{Memobase.} The primary purpose of Memobase is not to preserve arbitrary long-term context for an agent, but to build a persistently evolving structured profile for each user. Developers can explicitly define what user information the system should remember and organize that information into a clear \emph{topic and subtopic} structure. Compared with general-purpose memory layers based mainly on collections of discrete facts, this design more closely resembles a controllable memory database for individualized modeling. A second key feature of Memobase is its time-aware memory. In addition to structured profiles, it maintains user events and their timelines to support answering time-sensitive questions.

At the system implementation level, Memobase adopts a \texttt{buffer+asynchronous} flush writing mechanism. User conversations are first written to a buffer in blob form; when the buffer reaches a certain size, remains idle for a period of time, or when \texttt{flush()} is explicitly invoked after the session ends, the system converts the buffered content into long-term memory. Its context API can further package the user profile together with recent events into textual context that can be directly inserted into prompts. Compared with approaches that retain the full dialogue as long-term storage, this design reduces, to some extent, the risk of preserving raw sensitive text. At the same time, however, it also implies that private information may become sedimented into the profile and event timeline in the form of profiling and summarization.

\subsection{Training Pipeline and Optimization Strategy}
We adopt a two-stage training pipeline consisting of full-parameter SFT followed by reinforcement learning from reward feedback using GRPO. In both stages, we optimize the entire model rather than parameter-efficient adapters, so that the base model can be consistently adapted to the target instruction-following and policy-optimization objectives.

In the SFT stage, training is conducted with full-parameter updates under DeepSpeed ZeRO-3 optimization. We use the \texttt{qwen3\_nothink} template, disable thinking mode, and set the maximum sequence length to 4096 tokens. Data preprocessing uses 16 workers and dataloading uses 4 workers. Optimization is performed in \texttt{bf16} with a per-device batch size of 1 and gradient accumulation over 16 steps, yielding an effective batch expansion without exceeding device memory limits. We train for 1 epoch using Adam-style optimization with a learning rate of \(1.0\times10^{-5}\), cosine learning-rate decay, and a warmup ratio of 0.1. For monitoring and checkpointing, we log every 10 steps, save checkpoints every 20 steps, enable loss visualization, and overwrite the output directory for controlled reruns. A large DDP timeout is used to improve robustness in distributed training environments.

In the reinforcement learning stage, we further optimize the SFT model using GRPO implemented in \texttt{swift rlhf}. We set both \texttt{rlhf\_type} and \texttt{advantage\_estimator} to GRPO, and continue full-parameter training in \texttt{bfloat16}. To improve rollout efficiency, generation is served through a vLLM server backend. We disable dataset shuffling and training dataloader shuffling, and use left-side truncation so that the most recent context is preserved when samples exceed the context window. The maximum input length is 4096 tokens and the maximum completion length is 1536 tokens. Training uses a per-device batch size of 1, an evaluation batch size of 8, gradient accumulation over 8 steps, and 8 sampled generations per prompt. We set the learning rate to \(1\times10^{-6}\), warmup ratio to 0.01, maximum gradient norm to 1.0, PPO-style clipping parameter \(\epsilon=0.2\), and KL regularization coefficient \(\beta=0.001\). We train for 1 epochs with a single GRPO update iteration per batch, using DeepSpeed ZeRO-2 during this stage. For exploration, we use temperature \(=1.0\), top-\(p=0.9\), and top-\(k=50\), together with dynamic sampling and up to 3 resampling attempts. We evaluate every 20 steps, log every step, record generated completions and entropy statistics, save checkpoints every 20 steps, retain at most 3 checkpoints, and save model weights only.

Overall, this optimization strategy is designed to balance adaptation quality, training stability, and system efficiency. Full-parameter SFT provides a strong supervised initialization, while GRPO further improves the policy under reward-driven optimization. The use of long-context training, conservative learning rates in the RL stage, mixed-precision computation, DeepSpeed memory optimization, and vLLM-based rollout serving enables stable end-to-end training under limited GPU memory while maintaining reasonably efficient online sampling and checkpoint management.

\lstdefinelanguage{plaintext}{
    basicstyle=\ttfamily\small,
    showstringspaces=false,
    breaklines=true,
    frame=single,
    backgroundcolor=\color{gray!5},
}

\begin{figure*}[t]
\centering
\begin{lstlisting}[language=plaintext]
You are a memory retrieval assistant. Your task is to answer a question using only the retrieved conversation memories between a USER and an AI ASSISTANT.
# CONTEXT
You are given timestamped conversation memories from two participants: USER and ASSISTANT.
These memories come from previous conversations and may contain information needed to answer the question.
# INSTRUCTIONS
1. Carefully review all retrieved memories from both USER and ASSISTANT.
2. Use only the information explicitly stated in the memories.
3. Pay attention to timestamps to determine when events happened.
4. If multiple memories conflict, prioritize the most recent one.
5. If a memory contains relative time references (e.g., "last year", "two months ago"):
   - Convert them into an exact date, month, or year using the memory timestamp.
   - Example: If a memory dated **4 May 2022** says "went to India last year", the event happened in **2021**.
6. Always convert relative time expressions into specific dates or years in your reasoning.
7. Do not assume facts that are not present in the memories.
8. Treat USER and ASSISTANT only as speakers in the conversation. Do not confuse people mentioned inside memories with the speakers themselves.
9. The final answer must be **short**.
# REASONING PROCESS
Think step by step:
1. Identify memories relevant to the question.
2. Examine their timestamps and content.
3. Extract explicit facts about dates, locations, or events.
4. Convert relative time references to exact dates if necessary.
5. Select the most reliable evidence (prefer newer memories if conflicts exist).
6. Produce a concise answer that directly answers the question.
# MEMORY DATA
Memories from USER ({user_name}):
{user_memories}
# QUESTION
{question}
# ANSWER
\end{lstlisting}
\caption{Prompt template for answering MemPrivacy-Bench short-answer questions.}
\label{fig:prompt_for_qa}
\end{figure*}

\begin{figure*}[t]
\centering
\begin{lstlisting}[language=plaintext]
You are an assistant that selects the most appropriate answer for a user based on their query and known preferences.
Query:
{question}
Relevant User Memory:
{user_memories}
Candidate Answers:
{options_text}
Task:
Choose the answer that best fits the query while considering the user's preferences and past behavior from the provided memory.
Instructions:
- Carefully consider the user's memory when making your choice.
- You must strictly base your selection only on the provided memory and options; do not make assumptions, guesses, or introduce any information not explicitly supported by the memory.
- Select the single best option.
- Provide a short reason explaining why the selected option best matches the user's query and preferences.
Output Format:
Return your answer strictly in JSON format with the following fields:
{"answer": "<LETTER>", "reason": "<short explanation>"}
Rules:
- "answer" must be one of the option letters (e.g., A, B, C, D).
- The explanation should be concise (1~2 sentences).
- Do not output anything other than the JSON object.
\end{lstlisting}
\caption{Prompt template for answering PersonaMem-v2 multiple-choice questions.}
\label{fig:prompt_for_mcq}
\end{figure*}

\begin{figure*}[t]
\centering
\begin{lstlisting}[language=plaintext]
You are an **expert evaluator** for question-answering in an **AI memory system**.
Your task is to **strictly evaluate the accuracy** of the **Memory System Response** based **only** on the provided **Question** and **Reference Answer**.
* Determine whether the response is **"correct"**, **"partially_correct"**, or **"incorrect"**.
* **Do not use any external knowledge, assumptions, or subjective reasoning.**
* Your judgment must rely **exclusively** on the **Reference Answer** and the **Memory System Response**.
* Output your final decision **strictly in the required JSON format**.
# Evaluation Criteria
## 1. Answer Classification
### Correct
The **Memory System Response** is considered **correct** if:
* It **accurately answers the Question**.
* Its meaning is **semantically equivalent** to the **Reference Answer**.
* It **does not contradict** the Reference Answer.
* It **does not introduce unsupported or fabricated details**.
* **Synonyms, paraphrases, and reasonable summarizations** are acceptable.
### Partially Correct
The **Memory System Response** is considered **partially_correct** if:
* The response **contains some correct information from the Reference Answer**, but **does not fully cover all required elements**.
* The response is **incomplete**, but the information it provides is **consistent with the Reference Answer**.
* The response **does not contain contradictions** with the Reference Answer.
* The response **does not fabricate or invent unsupported facts**.
Typical cases include:
* The **Reference Answer contains multiple elements**, but the response **only includes some of them**.
* The response **captures the main idea but lacks important details** required for full correctness.
### Incorrect
The response is **incorrect** if **any** of the following conditions occur:
* It **contradicts** the Reference Answer.
* It **contains fabricated, unsupported, or invented information**.
* It **answers a different question** or is **irrelevant**.
* The response is **empty, meaningless, or non-informative**.
* The response **fails to include any correct information from the Reference Answer**.
## 2. Priority Rules (Conflict Handling)
1. **Contradictory or fabricated information always results in `incorrect`**, even if some parts are correct.
2. If the response **contains only a subset of the Reference Answer but remains fully consistent**, classify it as **`partially_correct`**.
3. A response is **`correct` only if it fully captures the meaning of the Reference Answer**.
\end{lstlisting}
\caption{Prompt template for GPT-5.2-Based grading of short-answer responses (1/2).}
\label{fig:prompt_for_judge_1}
\end{figure*}

\begin{figure*}[t]
\centering
\begin{lstlisting}[language=plaintext]
## 3. Detailed Guidelines and Tolerances
* **Equivalent expressions** of numbers, time, or units are acceptable, but the **numerical values themselves must match**.
* For **multi-element questions**:
  * **All elements present and correct = correct**
  * **Only some elements present (no contradictions) = partially_correct**
  * **Missing all key elements or introducing wrong elements = incorrect**
* If the Reference Answer is **"unknown / cannot be determined"**:
  * If the system provides a **specific factual claim**, it is **incorrect**.
  * If the system also answers **"unknown"** without speculation, it may be **correct**.
* The evaluation must rely **only** on the **Reference Answer** and the **Memory System Response**.
  **External context, world knowledge, or inference is not allowed.**
# Evaluation Input
**Question:** {question}
**Reference Answer:** {reference_answer}
**Memory System Response:** {response}
# Output Requirements
Provide the evaluation result **strictly** in the following JSON format.
* **Do not include any explanations or comments outside the JSON block.**
```json {"reason": "Provide a concise evaluation rationale", "judgment": "correct / partially_correct / incorrect"}```
\end{lstlisting}
\caption{Prompt template for GPT-5.2-Based grading of short-answer responses (2/2).}
\label{fig:prompt_for_judge_2}
\end{figure*}

\begin{figure*}[t]
\centering
\begin{lstlisting}[language=plaintext]
You are a professional "Data Security and Privacy Compliance Expert." Your core task is to review user-AI dialogues and identify sensitive privacy information contained within.
# Task
You need to analyze the input dialogue text, strictly following the [Privacy Level Standards (PL1-PL4)] defined below, extract all information belonging to PL2, PL3, and PL4, and output it in the specified JSON format.
You are given:
1. A `User's Real Name` field: the user's verified real full name, provided to assist you in determining whether a name appearing in the dialogue is the user's own name or a third party's name.
2. A `Current Input Dialogue` field: the dialogue content between the user and the AI that you must analyze, from which you should identify and extract all fragments containing PL2, PL3, or PL4 privacy information according to the standards defined below.
# Privacy Level Standards & Classification Rules (Knowledge Base)
## 1. Core Principles (Must Observe)
  - Extraction Scope: Only extract PL2 (Identifiable), PL3 (High Sensitivity), and PL4 (Confidential) information.
  - Exclusion Scope: Strictly forbid extracting PL1 (Low Sensitivity/Preferences) information. Preferences, habits, non-diagnostic emotions, and tone/style are not considered privacy information for extraction.
  - Public Information Exception: Public Information Exception: Publicly known global/national-level public figures, well-known institutions, or famous locations that are part of general knowledge, and are not linked to the user's personal identity, trajectory, or private context in the dialogue, do not need to be identified or extracted.
  - Conflict Resolution:
    - Once a high-level rule (e.g., PL4) is matched, categorize it immediately; do not downgrade.
    - When uncertain, follow the "higher rather than lower" principle (PL2 -> PL3 -> PL4).
    - PL1 vs. PL2+: If information describes a habit (PL1) but contains a specific location (PL2), the location information must be extracted.
## 2. Detailed Definitions & Categories
### [PL4: Confidential/Credentials/Critical Loss] (Highest Priority)
  - Definition: Any authentication, authorization, signing, or access control material that can be "directly reused/immediately executed," or key secrets that, if leaked, could immediately lead to account takeover, financial loss, system lateral movement, or mass data exfiltration.
  - Core Standard: Usable immediately upon acquisition, requiring no social engineering, directly leading to account takeover or financial loss.
  - Classification Rules:
    1. Auth/Account: Passwords, PINs, Security Questions & Answers, Verification Codes (SMS/Email/MFA), Session Tokens, Cookies (containing auth), OAuth Codes, Bank/Payment Card Security Codes (CVC, CVV, etc.), Backup Codes, Recovery Codes, SSO Tickets.
    2. Keys/Signatures: API Keys, AccessKeys, Secret Keys, Private Keys, Mnemonics, Seed Phrases, Database Connection Strings (containing credentials), Certificate Private Keys, Signing Keys, Encryption Keys, etc.
    3. System/Attack: Database strings, Admin portal URLs, Reproducible vulnerability details, Intranet entry points/Internal network segments, Bastion host info, CI keys, Cloud keys, Production configurations, etc.
\end{lstlisting}
\caption{Prompt template for privacy extraction (1/4).}
\label{fig:prompt_privacy_extraction_1}
\end{figure*}

\begin{figure*}[t]
\centering
\begin{lstlisting}[language=plaintext]
    4. Undisclosed Business Info: Undisclosed financials, M&A materials, Core roadmaps, Internal pricing, Client lists, Contract originals, Core implementations, Exploit details, Vulnerability PoCs, etc.
  - Standard Type Tags: Password, Verification Code, Token, Key, Private Key, Payment Security Code, Database Connection String, Vulnerability Details, Business Secret.
### [PL3: Highly Sensitive PII] (High Risk)
  - Definition: Information that, if leaked or illegally used, is expected to cause significant harm to personal safety/property, physical/mental health, reputation, or fair opportunity; or data belonging to generally sensitive categories.
  - Core Standard: High damage consequences. Even if it may not uniquely identify an identity on its own, it should be classified as PL3.
  - Classification Rules:
    1. Documents: ID Card Number, Passport Number, Social Security/Insurance Number, Document Photos/Scans, Driver's License Number, License Plate Number, etc.
    2. Financial: Bank/Payment Card Number, Basic Card Info (Opening Bank/Card Org/Type/Validity or Expiry Date, etc.), Account Info, Transaction Records/Bill Details, Salary/Income (Annual/Monthly income), Credit Reports (Credit Score/Points), Debt/Loan Info, Assets/Net Worth.
        - [Note] Transaction records/Bill details require judgment based on specific purpose and behavior. If it is just daily consumption behavior involving no exposure of personal privacy, do not classify (e.g., "Spent 86 yuan at the supermarket"). However, "Spent 1800 yuan for a checkup at a fertility clinic" or "Bank card ending in xxxx deducted 500 yuan" requires classification as they involve health and financial privacy respectively.
    3. Health: Medical Records/History/Hospital Visits/Surgery & Clinical Procedures, Diagnosis Results, Prescriptions, Specific Physiological Metrics (Blood Type/Blood Sugar/Blood Pressure/Lipids/Blood Oxygen, etc.), Specific Body Metrics (Height/Weight/BMI, etc.), Reproductive Health, Mental Illness/Therapy or Counseling Records (Note: Non-diagnostic emotional descriptions should be classified as PL1). Physiological and body metrics should only be classified as PL3 when specific values are given; qualitative descriptions should not be classified.
    4. Trajectory: Precise Location (Latitude/Longitude/Real-time positioning), Accommodation Records (Hotel Room Number, Check-in Time, etc.), Detailed Trajectory (Travel Itinerary, Train/Plane Ticket Info), Commute Routes, etc.
    5. Biometrics: Face, Fingerprint, Voiceprint, Iris features, etc.
    6. Communication Content: Raw Chat Logs, SMS/Email Content (not just contact info), Call Detail Records, etc.
    7. Sensitive Attributes: Ethnicity/Race/Tribe, Religious Beliefs, Political Views/Stance.
    8. Others: Minor Information (Under 14, Guardian info), Litigation/Arbitration/Penalty Records/Police Reports, etc.
  - Standard Type Tags: ID Number, Financial Account, Transaction Record, Assets/Income, Medical Health, Precise Location, Itinerary/Trajectory, Biometrics, Communication Content, Sensitive Identity, Judicial Record.
\end{lstlisting}
\caption{Prompt template for privacy extraction (2/4).}
\label{fig:prompt_privacy_extraction_2}
\end{figure*}

\begin{figure*}[t]
\centering
\begin{lstlisting}[language=plaintext]
### [PL2: Identifiable PII] (Basic Identification)
  - Definition: Information that, alone or combined with reasonably available information, can identify, locate, or stably trace a specific natural person.
  - Core Standard: Identifiable / Linkable / Traceable.
  - Classification Rules:
    1.  Direct Identifier: Real Name (Full Name), Specific Age, Specific Date of Birth, Gender, Mobile Number, Landline, Email Address, Detailed Address (Street/Doorplate level, Community/Building, Deliverable Address, etc.), Zip Code, Work Address.
    2.  Network Identifier: Account Username/Account ID/Platform UID/Device Account Name, Personal Homepage Link, Device Identifier, IP Address, Device ID, UserAgent, Reusable Cookies/Session Identifiers.
    3.  Strong Combination: Combinations that can lock onto a person like "Company + Job Title + Name", "School + Class + Name". Employer/Company Name, Job Title/Rank, School, and Class information appearing alone also need to be classified due to the potential for collection and combination.
    4.  Third-Party Identifiable Info: Personal information of Emergency Contacts/Relatives/Friends (Name, Phone, Email, Address, Relationship to the subject, etc.).
  - Standard Type Tags: Real Name, Phone Number, Email, Detailed Address, Account ID/Username, Network Identifier, Identity Background, Relationship Info.
### [PL1: Public/Low Sensitivity] (Negative Examples - DO NOT EXTRACT)
  - Definition: Unable to identify a specific individual; merely style, preferences, or habits.
  - Core Standard: Unidentifiable + Low Harm + Not High Sensitivity.
  - Classification Rules: Expression and interaction preferences, personality and emotional self-descriptions (non-diagnostic level), life rhythm and habit preferences, interest and content preferences, aesthetic and style preferences, motivation and goal preferences.
  - Typical Cases (Ignore this type of information):
    - "I like speaking in this tone" (Expression preference)
    - "I run at 6 am every morning" (General habit)
    - "I've been under a lot of pressure lately" (Non-diagnostic emotion)
    - "I like watching sci-fi movies" (Interest preference)
    - "I have a quick temper" (Personality self-description)
# Extraction Granularity & Boundary Principles
Core Principle: Only extract "Sensitive Entities" or "Minimum Sensitive Fact Fragments." Strictly forbid extracting full sentences, which would compromise the semantic integrity of the original dialogue.
1.  Remove Unnecessary Context:
    - Do not include introductory words (e.g., "My number is," "I live at," "The doctor said").
    - Do not include punctuation marks (unless part of an address or numerical value).
    - Example:
      - Original: "I live at Zhongguancun, Haidian District, Beijing" -> Extract: "Zhongguancun, Haidian District, Beijing" (Not the full sentence)
      - Original: "My password is 123456" -> Extract: "123456" (Not the full sentence)
2.  Maintain Semantic Integrity (For Descriptive Privacy):
    - For privacy that cannot be summarized in a single word (like transaction details, trajectories), extract the minimum phrase containing the core elements.
    - Example:
\end{lstlisting}
\caption{Prompt template for privacy extraction (3/4).}
\label{fig:prompt_privacy_extraction_3}
\end{figure*}

\begin{figure*}[t]
\centering
\begin{lstlisting}[language=plaintext]
      - Original: "I didn't feel well last night, so I spent 1800 yuan for a checkup at the fertility clinic" -> Extract: "spent 1800 yuan for a checkup at the fertility clinic" (If only "1800 yuan" is extracted, the transactional meaning is lost).
      - Original: "I have severe anxiety disorder" -> Extract: "severe anxiety disorder"
3.  Values Must Combine with Unit/Object:
    - Standalone numbers (e.g., "300") are generally not extracted unless they are phone numbers, ID numbers, or specific amounts matching PL2-PL4 rules.
    - For privacy involving amounts, extract the "Amount + Purpose" combination (if they appear together). *[Note]* Judgment must be based on the privacy level of the behavior/purpose. If the behavior meets PL2-PL4 rules, extract "Amount + Purpose"; otherwise, do not extract.
4.  Real Name Must Be the User's Own Full Name
    - Only the user's own full name qualifies as Real Name (PL2).
    - Use the provided `User's Real Name` field as the authoritative reference to determine whether a name in the dialogue belongs to the user. A name in the dialogue that matches or is a recognizable variant of the `User's Real Name` (e.g., with/without title, with/without middle name, different transliteration) should be treated as the user's own name. Names that do NOT match the `User's Real Name` should be treated as third-party names.
# Output Format (Requirements)
Please strictly follow the JSON format for output. Do not include Markdown code block markers (like ```json). Output the JSON array directly. If no PL2-PL4 information is found, output an empty array `[]`.
JSON Field Explanation:
  - `original_text`: Must directly copy the original text fragment from the dialogue without modification, masking, or summarization.
  - `privacy_type`: Select from the "Standard Type Tags" defined above; if an exact match is not possible, provide a corresponding type based on semantic judgment. The value must be in English.
  - `privacy_level`: Limited to `PL2`, `PL3`, `PL4`.
## Example (One-Shot)
Input Text:
User's Real Name: Zhang San
Current Input Dialogue: {"role": "user", "content": "Hello, my name is Zhang San, and my mobile number is 13800138000. I've been having insomnia recently, and the doctor diagnosed me with mild depression. Here is a photo of my prescription. Also, I just received a verification code 89757, please fill it in for me. By the way, I like spicy food and I speak quite directly."}
Output:
[{"original_text": "Zhang San", "privacy_type": "Real Name", "privacy_level": "PL2"},
 {"original_text": "13800138000", "privacy_type": "Phone Number", "privacy_level": "PL2"},
 {"original_text": "mild depression", "privacy_type": "Medical Health", "privacy_level": "PL3"},
 {"original_text": "89757", "privacy_type": "Verification Code", "privacy_level": "PL4"}]
(Note: PL1 information like "like spicy food" and "speak directly" was ignored)
# Input
User's Real Name: {real_name}
Current Input Dialogue:
\end{lstlisting}
\caption{Prompt template for privacy extraction (4/4).}
\label{fig:prompt_privacy_extraction_4}
\end{figure*}

\begin{longtable}{p{2cm} p{1.5cm} p{0.9cm} p{6.5cm} p{3.7cm}}
\caption{Illustrative examples of detailed privacy policy provisions.}
\label{tab:detailed_privacy_policy} \\
\toprule
\textbf{Primary Governance Principle} & \textbf{Secondary Theme} & \textbf{Privacy Level} & \textbf{Policy Provision} & \textbf{Handling Strategy} \\
\midrule
\endfirsthead

\toprule
\textbf{Primary Governance Principle} & \textbf{Secondary Theme} & \textbf{Privacy Level} & \textbf{Policy Provision} & \textbf{Handling Strategy} \\
\midrule
\endhead

\midrule
\multicolumn{5}{r}{\textit{Continued on the next page}} \\
\midrule
\endfoot

\bottomrule
\endlastfoot

Minimization \& Revocability
& Expression and Interaction Preferences
& 1
& \textbf{[Root]} Retain only preference summaries necessary for user experience; personally identifying details must not be stored. \newline
\textbf{[Model]} When personalization is applied, preferences must not be treated as factual identity attributes. \newline
\textbf{[User]} Provide user controls to enable or disable personalized memory and to clear memory with one click.
& Collection: record only when the user explicitly expresses a preference. \newline
Storage: store summarized representations rather than verbatim utterances. \newline
Retention: short-lived or user-configurable.
\\

Minimization \& Revocability
& General Habits
& 1
& \textbf{[Root]} Do not collect precise schedules; retain only coarse-grained tendencies. \newline
\textbf{[Developer]} Impose granularity limits on fields related to routines, diet, exercise, and similar habits.
& Collection: primarily based on user self-disclosure; avoid tracking-based collection. \newline
Storage: use coarse-grained enumerations.
\\

Minimization \& Revocability
& Capability and Modality Preferences
& 1
& \textbf{[Root]} Preferences regarding capability or mode of interaction must not be extrapolated into conclusions about qualification or competence.
& Use: only for interaction strategy. \newline
Sharing: individual profiles must not be exported externally.
\\

Minimization \& Revocability
& Interests and Aesthetic Preferences
& 1
& \textbf{[Root]} Do not retain feature combinations that could support sensitive inference; apply abstraction or fuzzing where necessary. \newline
\textbf{[User]} Allow users to inspect the list of memory entries.
& Storage: interest tags may be stored only if they remain visible to the user and deletable.
\\

Prudent Inference
& Personality and Emotion
& 1
& \textbf{[Root]} Emotional expressions must not be elevated into diagnostic judgments; no health profile may be created from such signals. \newline
\textbf{[Model]} Responses must include limiting language indicating that conclusions are based only on the current expression and do not constitute medical advice.
& Use: only for empathy and response style within the current dialogue. \newline
Retention: stress or emotion trajectories should not be stored long-term by default.
\\

Lawfulness \& Controlled Storage
& Direct Identifiers
& 2
& \textbf{[Root]} PL2 data must not enter general long-term memory by default. \newline
\textbf{[Developer]} If retention is strictly necessary, store only as structured fields with encryption, role-based access control, and audit logging. \newline
\textbf{[Model]} Avoid revealing full identifiers; masked presentation is preferred.
& Collection: only when strictly necessary to complete the service. \newline
Storage: controlled domain with field-level encryption and minimal retention.
\\

Lawfulness \& Controlled Storage
& Online Identifiers
& 2
& \textbf{[Root]} Usernames, UIDs, IP addresses, cookies, and similar identifiers must be handled according to their traceability; reusable session identifiers should be treated as higher risk. \newline
\textbf{[Developer]} Distinguish temporary session data from identifiers that support persistent tracking.
& Storage: avoid retention whenever possible; if retention is necessary, use a controlled domain with expiration policies.
\\

Lawfulness \& Minimization
& Identifiable Relationship and Background Combinations
& 2
& \textbf{[Model]} When combinations such as employer + title + city are detected, the user should be informed of the associated privacy risk and asked whether to proceed. \newline
\textbf{[Root]} Such combinations must not be used for external identity inference or external identity lookup.
& Use: limited to the current task only. \newline
Retention: not stored long-term by default.
\\

Security Protection \& Separate Consent
& Government Documents and Specific Identity Attributes
& 3
& \textbf{[Root]} Identity document numbers and document images must not be collected or stored in general-purpose systems; if exceptionally necessary, separate consent and stringent risk controls are required. \newline
\textbf{[Developer]} Document-related fields must be strongly encrypted, retained for the shortest possible period, and may require dual approval for access.
& Collection: collect only strictly necessary fields; verification outcomes are preferred over raw document content. \newline
Storage: controlled domain with strict auditing.
\\

Security Protection
& Financial and Credit Information
& 3
& \textbf{[Root]} Transaction details, bank card numbers, and credit records should not be collected by default; if proactively provided by the user, masking should be recommended first. \newline
\textbf{[Model]} Process only the minimally necessary fragments and return redacted content where needed.
& Storage: excluded from databases by default; if retention is necessary, use a controlled domain with field masking.
\\

Security Protection
& Health and Medical Information
& 3
& \textbf{[Root]} Diagnoses, prescriptions, and medical histories must not enter general-purpose memory and must not be used for profiling, advertising, or differential treatment. \newline
\textbf{[Model]} Responses must avoid diagnostic language and should recommend seeking professional medical care where appropriate.
& Use: limited to support within the current interaction; no health profile may be formed.
\\

Security Protection
& Precise Location and Trajectory
& 3
& \textbf{[Root]} Real-time location, trajectory point sequences, and lists of frequently visited places must not be retained long-term. \newline
\textbf{[Developer]} Apply upper bounds and automatic expiration to latitude--longitude and geofencing data.
& Storage: retain only city-level or region-level information if necessary. \newline
Retention: extremely short.
\\

Security Protection
& Biometrics
& 3
& \textbf{[Root]} Fingerprint, face, and voiceprint templates or embeddings must not enter dialogue memory or logs. \newline
\textbf{[Developer]} Biometric templates and matching results must be isolated across domains and protected by strong access control.
& Storage: dedicated secure domain with a minimal access surface.
\\

Security Protection
& Message and Accommodation Information
& 3
& \textbf{[Root]} Raw chat logs, email content, and accommodation details such as room numbers must not be stored by default; if pasted by the user, the user should be prompted to remove sensitive passages. \newline
\textbf{[Model]} Any quoted content in responses must be minimized.
& Use: process only for the current request; raw content must not persist across sessions.
\\

Highest-Level Protection
& Accounts and Authentication Credentials
& 4
& \textbf{[Root]} Passwords, OTPs, MFA backup codes, session tokens, and similar credentials must not be collected, stored, or echoed. \newline
\textbf{[Model]} Once such content appears, the system must immediately advise credential replacement and recommend a secure handling channel. \newline
\textbf{[Developer]} Detection and automatic redaction are required; such data must never be written to logs.
& Collection: blocked. \newline
Storage: zero retention. \newline
Handling: redact and alert in accordance with organizational procedures.
\\

Highest-Level Protection
& Keys and Signing Materials
& 4
& \textbf{[Root]} API keys, private keys, mnemonic phrases, certificate private keys, and similar secrets must never enter model context, memory, logs, or training data. \newline
\textbf{[Developer]} Secrets must never be stored; enforce key rotation workflows and least-privilege access.
& Handling: immediately recommend revocation and rotation.
\\

Highest-Level Protection
& Enterprise Secrets and Non-public Business Data
& 4
& \textbf{[Root]} Non-public financials, M\&A materials, roadmaps, and reusable implementation details of core algorithms must not enter general-purpose dialogue systems or external toolchains by default. \newline
\textbf{[Developer]} Internal materials must be classified, access-isolated, and protected with watermarking and audit mechanisms for outbound sharing.
& Storage: dedicated internal domain with minimal visibility.
\\

\end{longtable}

\begin{algorithm}[h]
\renewcommand{\algorithmicrequire}{\textbf{Input:}}
\renewcommand{\algorithmicensure}{\textbf{Output:}}
\caption{MemPrivacy: A Closed-Loop Framework for Memory Privacy Protection}
\label{alg:memprivacy_framework}
\begin{algorithmic}[1]
\REQUIRE Raw user input $X$, local privacy extractor $f_{\theta}$, privacy protection threshold $\lambda \in \{\mathrm{PL1},\mathrm{PL2},\mathrm{PL3},\mathrm{PL4}\}$, local secure mapping database $\mathcal{D}$, cloud agent $C$, cloud-side sanitized memory state $M_{\mathrm{safe}}$
\ENSURE Restored personalized response $\hat{Y}$, updated local mapping database $\mathcal{D}'$

\STATE \textbf{Stage 1: Local Uplink Desensitization}
\STATE Detect privacy spans with structured attributes:
\[
\mathcal{P} \leftarrow f_{\theta}(X)=\{(s_i, l_i, t_i)\}_{i=1}^{N},
\]
where $s_i$ is the privacy span, $l_i$ is the privacy level, and $t_i$ is the privacy type
\STATE Initialize sanitized input $X_{\mathrm{safe}} \leftarrow X$

\FOR{each detected item $(s_i,l_i,t_i)\in\mathcal{P}$}
    \IF{$l_i \geq \lambda$}
        \IF{there exists a consistent mapping $(p_i \leftrightarrow s_i)$ in $\mathcal{D}$}
            \STATE Reuse placeholder $p_i$
        \ELSE
            \STATE $p_i \leftarrow \text{InstantiatePlaceholder}(t_i,\mathcal{D})$
            \STATE Store mapping $(p_i \leftrightarrow s_i, l_i, t_i)$ into $\mathcal{D}$
        \ENDIF
        \STATE Replace span $s_i$ in $X_{\mathrm{safe}}$ with $p_i$
    \ENDIF
\ENDFOR

\STATE \textbf{Stage 2: Cloud Processing}
\STATE Send $X_{\mathrm{safe}}$ to the cloud agent
\STATE Obtain cloud response based on $X_{\mathrm{safe}}$ and $M_{\mathrm{safe}}$:
\[
Y_{\mathrm{safe}} \leftarrow C(X_{\mathrm{safe}}, M_{\mathrm{safe}})
\]
\STATE Update cloud-side memory only with sanitized content:
\[
M_{\mathrm{safe}}' \leftarrow \text{UpdateMemory}(M_{\mathrm{safe}}, X_{\mathrm{safe}}, Y_{\mathrm{safe}})
\]

\STATE \textbf{Stage 3: Local Downlink Restoration}
\STATE Initialize restored response $\hat{Y} \leftarrow Y_{\mathrm{safe}}$
\FOR{each placeholder $p$ appearing in $\hat{Y}$}
    \IF{$p$ exists in local database $\mathcal{D}$}
        \STATE Replace $p$ with its original value $\mathcal{D}[p]$
    \ENDIF
\ENDFOR

\STATE \textbf{return} $\hat{Y}, \mathcal{D}'$
\end{algorithmic}
\end{algorithm}

\begin{table*}[t]
\centering
\caption{Hierarchical privacy level definitions and default handling policies.}
\label{tab:privacy_governance_principles}
\small
\setlength{\tabcolsep}{5pt}
\renewcommand{\arraystretch}{1.25}
\begin{tabularx}{\textwidth}{p{1.5cm} p{1.5cm} p{1.5cm} p{3cm} p{1.5cm} p{2.7cm} X}
\toprule
\textbf{Primary Governance Principle} & \textbf{Secondary Theme} & \textbf{Privacy Level} & \textbf{Decision Criteria} & \textbf{Long-Term Text Memory} & \textbf{Default Storage and Access Policy} & \textbf{Default Model Behavior} \\
\midrule
Lawfulness, legitimacy, necessity, and data minimization
& Classification and identifiability
& PL1: Low-Sensitivity Personalization
& Preferences, habits, or profiling cues; insufficient for direct identification; outside high-risk domains, though still potentially linkable or inferable
& Permitted, subject to minimization and revocability
& Preference summaries and vectorized features may be stored; avoid retaining directly traceable original utterances; provide one-click deletion
& May be used for personalized interaction; avoid reifying inferred profiles as facts
\\

Lawfulness, legitimacy, necessity, and data minimization
& Classification and identifiability
& PL2: Identifiable Personal Information
& Information that can identify an individual, either alone or in combination, including direct identifiers and stable indirect identifiers
& Disallowed by default
& If strictly required for business purposes, store only in controlled systems with encryption, access control, and auditing; RAG should retain pointers rather than raw text
& Avoid repetition or verbatim exposure in dialogue; when necessary, prompt for user confirmation and apply de-identification
\\

Security safeguards and risk control
& High-risk sensitive domains
& PL3: Highly Sensitive Personal Information
& Disclosure may materially harm personal safety, property, reputation, or health, or result in discriminatory treatment; typically requires a stronger legal basis
& Not permitted
& Only under compelling necessity and a clear legal basis may such data enter controlled storage; enforce strict minimization, field-level encryption, and robust auditing
& Do not collect by default; if provided by the user, prioritize masking key fields and warning about the associated risks
\\

Security safeguards and highest-level protection
& Confidential information, credentials, directly actionable high-loss data
& PL4: Confidential Information, Credentials
& Once disclosed, can directly enable login, fund transfer, signature authorization, lateral movement, or expose undisclosed core business secrets
& Absolutely prohibited
& Prohibited from databases, memory, and logs; once detected, it must be redacted or blocked immediately; secrets must never enter model context
& Must refuse collection, storage, or disclosure; advise credential rotation and escalation through secure incident-response procedures
\\
\bottomrule
\end{tabularx}
\end{table*}


\end{document}